\documentclass{nature}

\usepackage{amssymb}
\usepackage{amsmath}
\usepackage{color,graphicx}
\usepackage{caption}
\usepackage{times}
\usepackage{float}
\usepackage[colorinlistoftodos]{todonotes}
\usepackage{marginnote}

\title{Underground test of gravity-related wave function collapse}

\author{Sandro Donadi,$^{1,^\dagger}$ Kristian Piscicchia,$^{2,3,^\star}$ Catalina Curceanu,$^{3,2}$ \\
Lajos Di\'osi,$^{4}$ Matthias Laubenstein,$^{5}$ Angelo Bassi,$^{6,1,\ast}$}

\makeatletter
\let\saved@includegraphics\includegraphics
\AtBeginDocument{\let\includegraphics\saved@includegraphics}
\renewenvironment*{figure}{\@float{figure}}{\end@float}
\makeatother

\begin{document}

\maketitle

\begin{affiliations}
\item Istituto Nazionale di Fisica Nucleare, Trieste Section, Via Valerio 2, 34127, Trieste, Italy.
\item Centro Fermi - Museo Storico della Fisica e Centro Studi e Ricerche ``Enrico Fermi'', Piazza del Viminale 1, 00184 Rome, Italy.
\item INFN, Laboratori Nazionali di Frascati, Via Enrico Fermi 40, 00044 Frascati, Italy.
\item Wigner Research Centre for Physics, H-1525 Budapest 114 , P.O.Box 49, Hungary.
\item INFN, Laboratori Nazionali del Gran Sasso, Via G. Acitelli 22, 67100 Assergi, Italy.
\item Department of Physics, University of Trieste, Strada Costiera 11, 34151 Trieste, Italy.\\
\normalsize{Email: $^\dagger$sandro.donadi@ts.infn.it, $^\star$kristian.piscicchia@cref.it, $^\ast$abassi@units.it}
\end{affiliations}

\begin{abstract}
Roger Penrose proposed that a spatial quantum superposition collapses as a back-reaction from spacetime, which is curved in different ways by each branch of the superposition. In this sense, one speaks of gravity-related wave function collapse. He also provided a heuristic formula to compute the decay time of the superposition --- similar to that suggested earlier by Lajos Di\'osi, hence the name Di\'osi-Penrose model. The collapse depends on the effective size of the mass density of particles in the superposition, and is random: this randomness shows up as a diffusion of the particles' motion, resulting, if charged, in the emission of radiation. Here, we compute the radiation emission rate, which is faint but detectable. We then report the results of a dedicated experiment at the Gran Sasso underground laboratory to measure this radiation emission rate. Our result sets a lower bound on the effective size of the mass density of nuclei, which is about three orders of magnitude larger than previous bounds. This rules out the natural parameter-free version of the Di\'osi-Penrose model.
\end{abstract}

\section*{\Large Main text}
Quantum Mechanics beautifully accounts for the behaviour of microscopic systems while, in an equally beautiful but radically different way, Classical Mechanics accounts for the behaviour of macroscopic objects. The reason why the quantum properties of microscopic systems---most notably, the possibility of being in the superposition of different states at once---do not seem to carry over to larger objects, has been the subject of a debate which is as old as the quantum theory itself, as exemplified by Schr\"odinger's cat paradox~\cite{schrodinger1935}.

It has been conjectured that the superposition principle, the building block of quantum theory,  progressively breaks down when atoms glue together to form larger systems \cite{leggett1980macroscopic, weinberg1989precision, bell2004speakable, ghirardi1986unified, adler2004quantum, weinberg2012collapse}. 
The reason is that the postulate of wave function collapse introduced by von Neumann, and now part of the standard mathematical formulation of the theory, according to which the quantum state of a system suddenly collapses at the end of a measurement process, though being very effective in describing what happens in measurements, clearly has a phenomenological flavour. There is no reason to believe that measurements are so special to temporarily suspend the quantum dynamics given by the Schr\"odinger equation and replace it with a completely different one.  More realistically, if collapses occur at all, they are part of the dynamics: in some cases, they are weak and can be neglected; in some other cases, such as in measurements, they become strong and rapidly change the state of a system. 
Decades of research in this direction has produced well-defined models accounting for the collapse of the wave function and the breakdown of the quantum superposition principle for larger systems~\cite{ghirardi1986unified, ghirardi1990markov, bassi2003dynamical, bassi2013models}, and now the rapid technological development has opened the possibility of testing them~\cite{Arndt:2014aa}. One question is left open: what triggers the collapse of the wave function?

In his lectures on gravitation, Feynman discusses how a breakdown of the quantum superposition principle at a macroscopic scale leaves open the possibility that gravity might not be quantized~\cite{feynman2018feynman}. Along this line of thinking, Penrose (and Di\'osi, independently) suggested that gravity, whose effects are negligible at the level of atoms and molecules, but increase significantly at the level of macroscopic objects, could be the source of the wave function collapse: ``{\it My own point of view is that as soon as a `significant' amount of space-time curvature is introduced, the rules of quantum linear superposition  must fail}"~\cite{penrose1990emperor}. When a system is in a spatial quantum superposition, a corresponding superposition of two different space-times  is generated. Penrose then gives arguments~\cite{penrose1996gravity, penrose2014gravitization, howl2019exploring} as to why nature ``dislikes'' and, therefore, suppresses superpositions of different space-times; the more massive the system in the superposition, the larger the difference in the two space-times and the faster the wave-function collapse. 

\begin{figure*}[h]
\centering
\includegraphics[width=15cm]{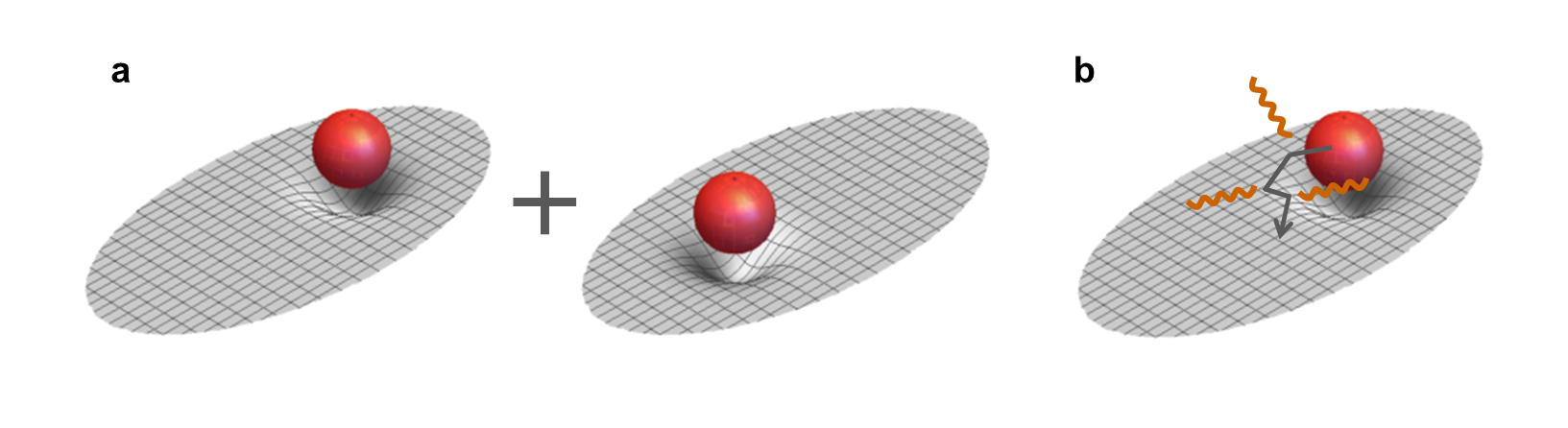}
\caption{\textbf{The Di\'osi-Penrose (DP) model of gravity-related wave function collapse.}
{\bf a.}  According to quantum gravity, a spatial quantum superposition of a system generates a superposition of different space-time curvatures, corresponding to the possible different locations of the system. Penrose argues that a superposition of different space-times is unstable and decays in time, making also the system’s wave function collapse. He provides an estimate for the time of collapse as given in Eq. (1), which is the faster the larger the system, similar to that suggested earlier by Di\'osi.
{\bf b.} The master equation of the DP model (Eq. (3)) predicts not only the collapse of the wave function, but also an omnipresent Brownian-like diffusion for each constituent of the system. When the constituents are charged (protons and electrons), the diffusion comes with the emission of radiation, with a spectrum that depends on the configuration of the system. This is given by Eq. (4) in the range $\Delta E = (10 \div 10^5) \mbox{keV}$ of photon energies. The predicted radiation emission is faint but potentially detectable by an experiment performed in a very low-noise environment. We performed such an experiment to rule out the original parameter-free version of the DP model.}
\label{superposition}
\end{figure*}
\clearpage

Even without proposing a detailed mathematical model, Penrose provides a formula which estimates, in non-relativistic and weak gravitational field limits, the expected time $\tau_{\text{\tiny DP}}$ of the collapse of a quantum superposition~\cite{penrose1996gravity}: 
\begin{equation}
\tau_{\text{\tiny DP}}=\frac{\hbar}{\Delta E_{\text{\tiny DP}}},\label{eq:tau}
\end{equation}
where $\Delta E_{\text{\tiny DP}}$ measures how large, in gravitational terms, the superposition is. Given a system with mass density $\mu(\boldsymbol{r})$, in the simple case of the center-of-mass being in a superposition of two states displaced by a distance $\boldsymbol{d}$:
\begin{equation} 
\label{eq:De Pen0}
\Delta E_{\text{\tiny DP}}(\boldsymbol{d}) =  - 8\pi G\int d\boldsymbol{r}\int d\boldsymbol{r}'\frac{\mu(\boldsymbol{r})\left[\mu(\boldsymbol{r}'+\boldsymbol{d})-\mu(\boldsymbol{r}')\right]}{|\boldsymbol{r}-\boldsymbol{r}'|}.
\end{equation}
Eqs.~\eqref{eq:tau} and \eqref{eq:De Pen0}, which are valid in the Newtonian limit, were previously proposed by Di\'osi \cite{diosi1987universal, diosi1989models}, following a different approach.
For a point-like mass density $\mu(\boldsymbol{r})=m\delta(\boldsymbol{r}-\boldsymbol{r}_0)$, Eq.~\eqref{eq:De Pen0} diverges because of the $1/r$ factor, leading to an instantaneous collapse, which is clearly wrong. To avoid this problem, one has to smear the mass density. This is implemented in different ways by Di\'osi and Penrose. Di\'osi suggests to introduce a new phenomenological parameter, measuring the  spatial resolution of the mass density~\cite{ghirardiDP, diosi2013gravity}; Penrose instead suggests that the mass density of a particle is given by $\mu(\boldsymbol{r})=m|\psi(\boldsymbol{r},t)|^2$~\cite{penrose2014gravitization}, where $\psi(\boldsymbol{r},t)$ is a stationary solution  of the
Schr\"odinger-Newton equation~\cite{DIOSI_SN, bahrami2014schrodinger}. For either choice, we will call $R_0$ the size of the particle's mass density.

A direct test of Eq.~\eqref{eq:tau} requires creating a large  superposition of a massive  system, to guarantee that $\tau_{\text{\tiny DP}}$ is  short enough for the collapse to become effective before any kind of external noise disrupts the measurement (see \cite{salart2008spacelike} for an alternative approach). One of the first proposals in this direction was put forward by Penrose himself and collaborators \cite{marshall2003towards}, who suggested a setup for creating a spatial superposition of a mirror of mass $\sim 10^{-12}$ Kg that, according to Eq.~(\ref{eq:De Pen0}), has a decay time of order $\tau_{\text{\tiny DP}}\simeq 0.002 \div 0.013$ s (see Supplementary Information), which is competitive with standard decoherence times. 
The major difficulty in implementing this and similar proposals consists in creating a superposition of a relatively large mass and keep it stable for times comparable to $\tau_{\text{\tiny DP}}$. To give some examples, the largest spatial superposition so far achieved~\cite{kovachy2015quantum} is of about  $0.5$ m,  but the systems involved are Rb atoms (mass = $1.42\times10^{-25}$ Kg), which are too light. In matter-wave interferometry with macromolecules~\cite{fein2019quantum}, states are delocalized over distances of hundreds nm, and masses beyond 25 kDa ($\sim10^{-23}$ Kg), still not enough. By manipulating phononic states \cite{lee2011entangling}, collective superpositions of estimated $10^{16}$ carbon atoms (mass $\sim 10^{-10}$ Kg) are created over distances of $10^{-11}$ m, but the life-time of phonons is of order $\sim 10^{-12}$ s, which is too short. These numbers show that keeping the superposition time, distance and mass large enough poses still huge technological challenges. Research towards creating larger and larger superpositions is very active~\cite{chan2011laser, teufel2011sideband, wollman2015quantum, jain2016direct, hong2017hanbury, vovrosh2017parametric, riedinger2018remote}, but further development is needed to reach the required sensitivity.

Here we show how to test gravitational-related collapse in an indirect way, by exploiting an unavoidable side effect of the collapse: a Brownian-like diffusion of the system in space. The reason is the following. Although Penrose restrains from proposing any detailed dynamics for the collapse, as suggested in~\cite{penrose1996gravity,penrose2014gravitization} and used explicitly in~\cite{howl2019exploring},    the simplest assumption is that the collapse is Poissonian, as for particle decay.  This minimal requirement, together with the collapse time given in Eqs. (\ref{eq:tau}) and (\ref{eq:De Pen0}), implies  the following Lindblad dynamics for the statistical operator $\rho(t)$ describing the state of the system (see Supplementary Information):  
\begin{equation}
\label{eq:dfgdfgd}
\frac{d\rho(t)}{dt}=-\frac{i}{\hbar}\left[H,\rho(t)\right]-\frac{4\pi G}{\hbar}\int d\boldsymbol{x}\int d\boldsymbol{y}\frac{1}{|\boldsymbol{x}-\boldsymbol{y}|}\left[\hat{M}(\boldsymbol{y}),\left[\hat{M}(\boldsymbol{x}),\rho(t)\right]\right].
\end{equation}
which is equivalent to the master equation derived in~\cite{diosi1987universal, diosi1989models}. The first term describes the standard quantum evolution while the second term accounts for the gravity-related collapse. In Eq. (\ref{eq:dfgdfgd}) $H$ is the system's Hamiltonian and $\hat{M}(\boldsymbol{x})=\sum_{n}\mu_{n}(\boldsymbol{x},\hat{\boldsymbol{x}}_{n})$ gives the total mass density, with $\mu_{n}(\boldsymbol{x},\hat{\boldsymbol{x}}_{n})$ the mass density of the $n$-th particle, centered around $\hat{\boldsymbol{x}}_{n}$. Taking for example a free particle with momentum operator $\hat{\boldsymbol{p}}$, the contribution of the second term to the average momentum $\langle \boldsymbol{p} \rangle \equiv \text{Tr}[\hat{\boldsymbol{p}} \rho]$ is zero, while the contribution to the average square momentum $\langle \boldsymbol{p}^2 \rangle$ increases in time. This is diffusion.

This diffusion causes a progressive heating of the system \cite{ghirardiDP}, specifically a steady temperature increase. Assuming a  mass distribution of the nuclei  with an effective size $R_0 \sim 10^{-15}$ m, the heating rate for a gas of non-interacting particles amounts to: $dT(t)/dt =4\sqrt{\pi}m_{0}G\hbar/3k_{\textrm{\tiny B}}R_{0}^{3}\sim10^{-4}\textrm{ K/s}$ ($k_{\textrm{\tiny B}}$ is Boltzmann's constant and $m_0$ the nucleon mass), which is in contradiction with experimental evidence~\cite{bahrami2014role}. 
The value $R_0 \sim 10^{-14}$ m is also excluded by gravitational wave detection experiments \cite{helou2017lisa}. However, both results do not include the possibility of dissipative effects, which are always associated to fluctuations, which may lead to equilibrium instead of a steady growth in temperature. 

Whether at thermal equilibrium or not, particles will keep fluctuating under the collapse dynamics. Since matter is made of charged particles, this process makes them constantly radiate. Therefore, a detection of the collapse-induced radiation emission is a more robust test of the model (cf.~\cite{diosi1993calculation}) , even in presence of dissipative effects.

Starting  from Eq.~\eqref{eq:dfgdfgd}, we computed the radiation emission rate, i.e. the number of photons emitted per unit time and unit frequency, integrated over all directions, in the range $\lambda \in (10^{-5} \div 10^{-1})$ nm, corresponding to energies $E \in (10 \div 10^{5})$ keV.
 The reason for choosing this range can be understood in terms of a semi-classical picture: each time a collapse occurs, particles  are slightly and randomly moved. This random motion makes them emit radiation, if charged. When their separation is smaller than $\lambda$, they emit as a single object with charge equal to the total charge, which can be zero for opposite charges as for an atom. Opposite to this, when  their separation is larger than $\lambda$, they emit independently. Therefore, in order to maximize the emission rate, electrons and nuclei should be independent ($\lambda <$ atomic radius), while protons in the same nucleus should behave coherently ($\lambda >$ nuclear radius). This is achieved by considering the emission of photons with wavelength in the range mentioned above. In this range, the coherent emission of protons contributes with a term proportional to $(Ne)^2$ ($N$ is the atomic number), while electrons contribute incoherently with a weaker term proportional to $Ne^2$. For this reason, and also because in the range of energies considered in our experiment the electrons are relativistic, while our derivation is not, to be conservative we will neglect the contribution of the electrons in the emission rate.

The photon emission rate is discussed in the Section Methods and derived in the Supplementary Information. The calculation is lengthy. In a nutshell, starting from  Eq.~(3), we compute the expectation value of the photon number operator at time $t$, i.e. $\langle a_{\mathbf{k}\mu}^{\dagger}a_{\mathbf{k}\mu}\rangle_{t}$, to the first perturbative order. By taking the time derivative, summing over the photon's polarizations $\mu$ and integrating over all the directions of the emitted photon, we eventually obtain:
\begin{equation}
\frac{d\Gamma_{t}}{d\omega}=\frac{2}{3}\frac{Ge^{2} N^{2} N_a}{\pi^{3/2}\varepsilon_{0}c^{3}R_{0}^{3}\omega},
\label{eq:rate}
\end{equation}
where $G, e, \varepsilon_0$ and $c$ are constants of nature with the usual meaning and $N_a$ is the total number of atoms. We leave $R_{0}$  as a free parameter to be bounded by experiments.  
Clearly, the number of emitted photons increases with the size ($N_a$) of the system, as there are more protons affected by the noise. The factor $N^2$ accounts for the quadratic dependence on the atomic number, which significantly increases the predicted effect.   

We performed, for the first time, a dedicated experiment to test this model of gravity-related collapse by measuring the spontaneous  radiation emission rate from a Germanium crystal and the surrounding materials in the experimental apparatus. The strong point of the experiment is that there was no need to create a spatial superposition, since according to Eq.~(\ref{eq:dfgdfgd}) the collapse induced diffusion and the associated photons emission occur for any state, also for localized states of the system.
The experiment was carried out in the low background environment of the Gran Sasso underground National Laboratory (LNGS) of INFN. 
The Gran Sasso Laboratory is particularly suitable for high sensitivity measurements of extremely low rate physical processes, since it is characterized by a rock overburden corresponding to a minimum thickness of 3100 m w.e. (meters of water equivalent). The environmental emissions are generated by the rock radioactivity and the residual cosmic muon flux. Given that the cosmic radiation flux is reduced by almost six orders of magnitude, the main background source in the LNGS consists of $\gamma$ radiation produced by long lived $\gamma$ emitting primordial isotopes and their decay products. They are part of the rocks of the Gran Sasso mountains and
the concrete used to stabilize the cavity.

\begin{figure*}[h]
\centering
\includegraphics[width=0.8\textwidth]{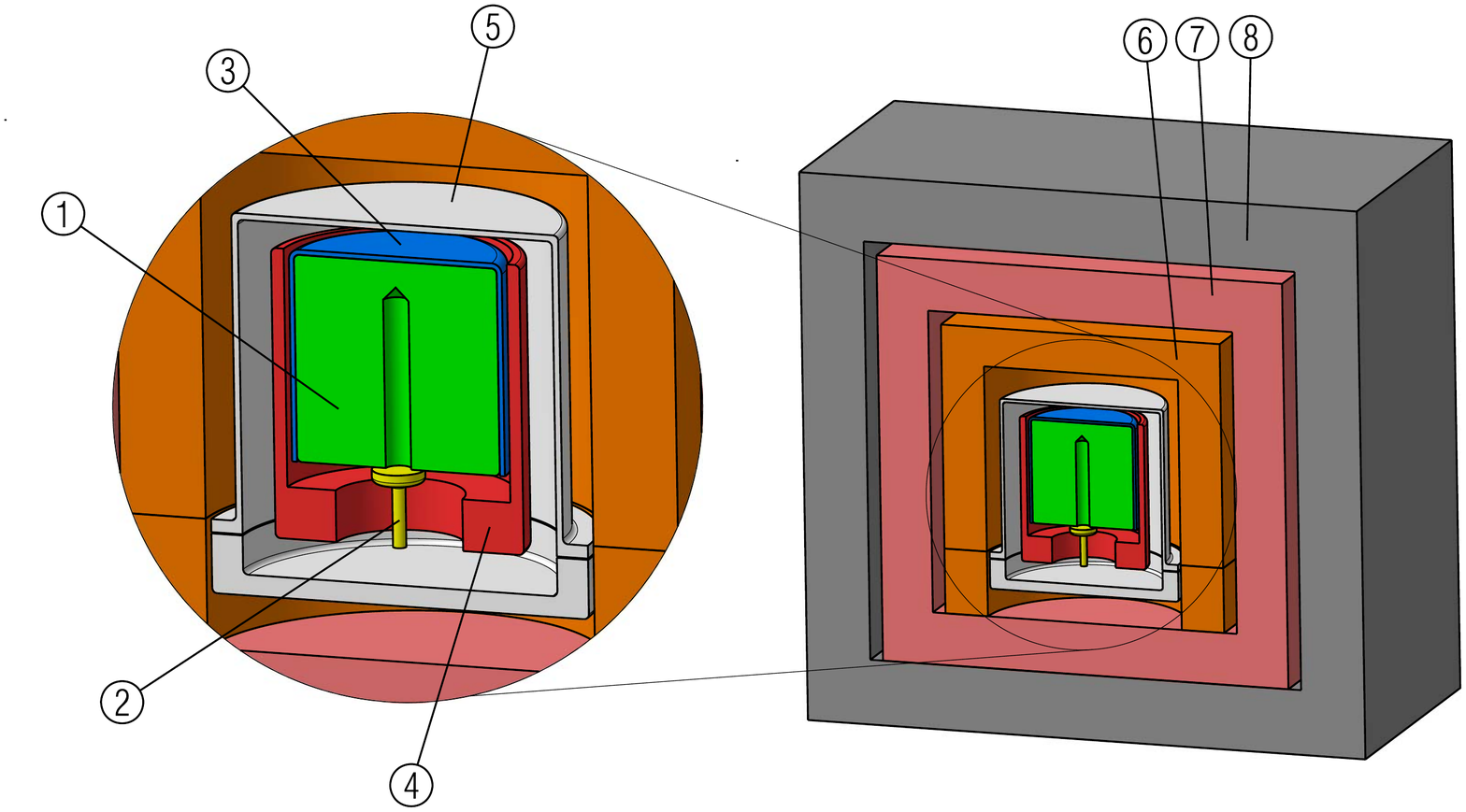}
\caption{\textbf{Schematic representation of the experimental setup.} The experimental apparatus is based on a coaxial p-type high purity germanium detector, with the dimensions of 8.0 cm diameter and 8.0 cm length, the active volume is 375 cm$^3$. The detector is shielded by layers of electrolitic copper and pure lead. The inner part of the apparatus consists of the following main elements: 1 - Ge crystal, 2 - Electric contact, 3 - Plastic insulator, 4 - Copper cup, 5 - Copper end-cup, 6 - Copper block and plate, 7 - Inner Copper shield, 8 - Lead shield. In order to minimize the radon contamination an air tight steel casing (not shown) encloses the shield and is continuously flushed with boil-off nitrogen from a liquid nitrogen storage tank.}
\label{pic_setup}
\end{figure*}

The setup consisted of a coaxial p-type High Purity Germanium (HPGe) detector surrounded by a complex shielding structure with the outer part made of pure lead and the inner part made of electrolytic copper. 
The Germanium crystal is characterized by a diameter of 8.0 cm and a length of 8.0 cm, with an inactive layer of lithium-doped germanium of 0.075 mm all around the crystal. The active germanium volume of the detector is 375 cm$^3$. The outer part of the passive shielding of the HPGe detector consists of lead (30 cm from the bottom and 25 cm from the sides). The inner layer of the shielding (5 cm) is composed of electrolytic copper. The sample chamber has a volume of about 15 l (($250 \times 250 \times 240$) mm$^3$). The shield together with the cryostat are enclosed in an air tight steel housing of 1 mm thickness, which is continuously flushed with boil-off nitrogen from a liquid nitrogen storage tank, in order to reduce the contact with external air (and thus radon) to a minimum. 
The experimental setup is schematically shown in Fig.~\ref{pic_setup} (see also~\cite{neder2000low,heusser2006low}).
The data acquisition system is a Lynx Digital Signal Analyzer controlled via personal computer software GENIE 2000, both from Canberra-Mirion.
In this measurement,  the sample placed around the detector was 62 kg of electropolished oxygen free high conductivity copper in Marinelli geometry. 

\clearpage

\begin{figure*}[h]
\centerline{%
\includegraphics[width=10cm]{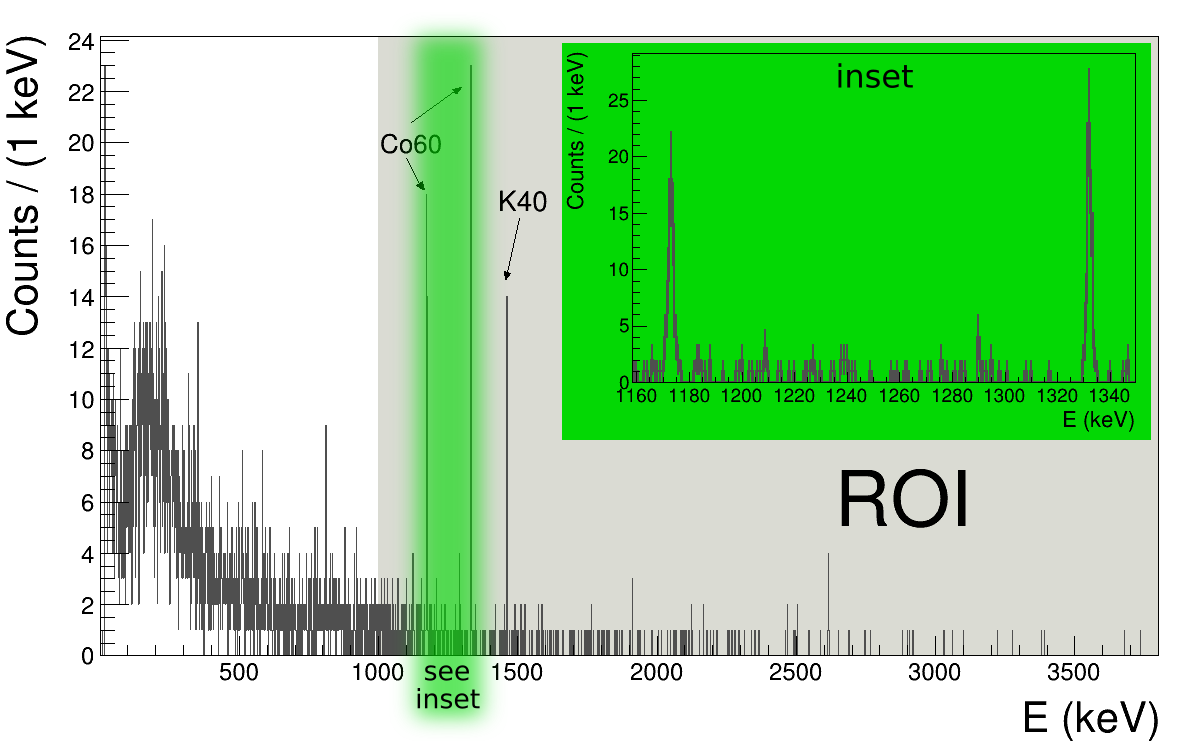}}
\caption{\textbf{Measured radiation spectrum.} The measured emission spectrum, corresponding to a data taking period of about 62 days, is represented as a dark grey histogram. The natural binning of 1 keV is used; the bin contents are shown without error bars, in order to appreciate the relative intensities of the residual radionuclides emission lines. The $^{60}$Co and $^{40}$K lines are also indicated. The inset zooms in the region of the $^{60}$Co lines (which is highlighted in the total plot by the shadowed green area); here the error bars are shown and represent one standard deviation.
The grey area shows the Region Of Interest (ROI) which is defined as $\Delta E = (1000 \div 3800)$ keV.}
\label{measured1}
\end{figure*}

The measured emission spectrum, corresponding to a data taking period of about 62 days (August 2014 and August 2015), is shown in Fig. \ref{measured1}, where emission lines generated by residual radionuclides present in the setup materials are also visible. 
In particular, the region of the $^{60}$Co lines (corresponding to the shadowed green area highlighted in the total plot) is enlarged in the inset.

Data analysis was carried out to extract the probability distribution function ($pdf$) of the $R_0$ parameter of the model. The novelty, with respect to previous investigations~\cite{fu1997spontaneous,piscicchia2017csl}, is not only the dedicated experiment but also an accurate Monte Carlo (MC) characterisation, with a validated MC code based on the GEANT4 software library, of the experimental setup, which allowed to compute the background originated from known sources, determining the contribution of each component of the setup; 
the background simulation is described in greater detail in the Section Methods. The residual spectrum was
then compared with the theoretical prediction for the collapse-induced radiation, to extract a bound
on $R_0$.

\begin{figure}[H]
\centering
\includegraphics[width=10cm]{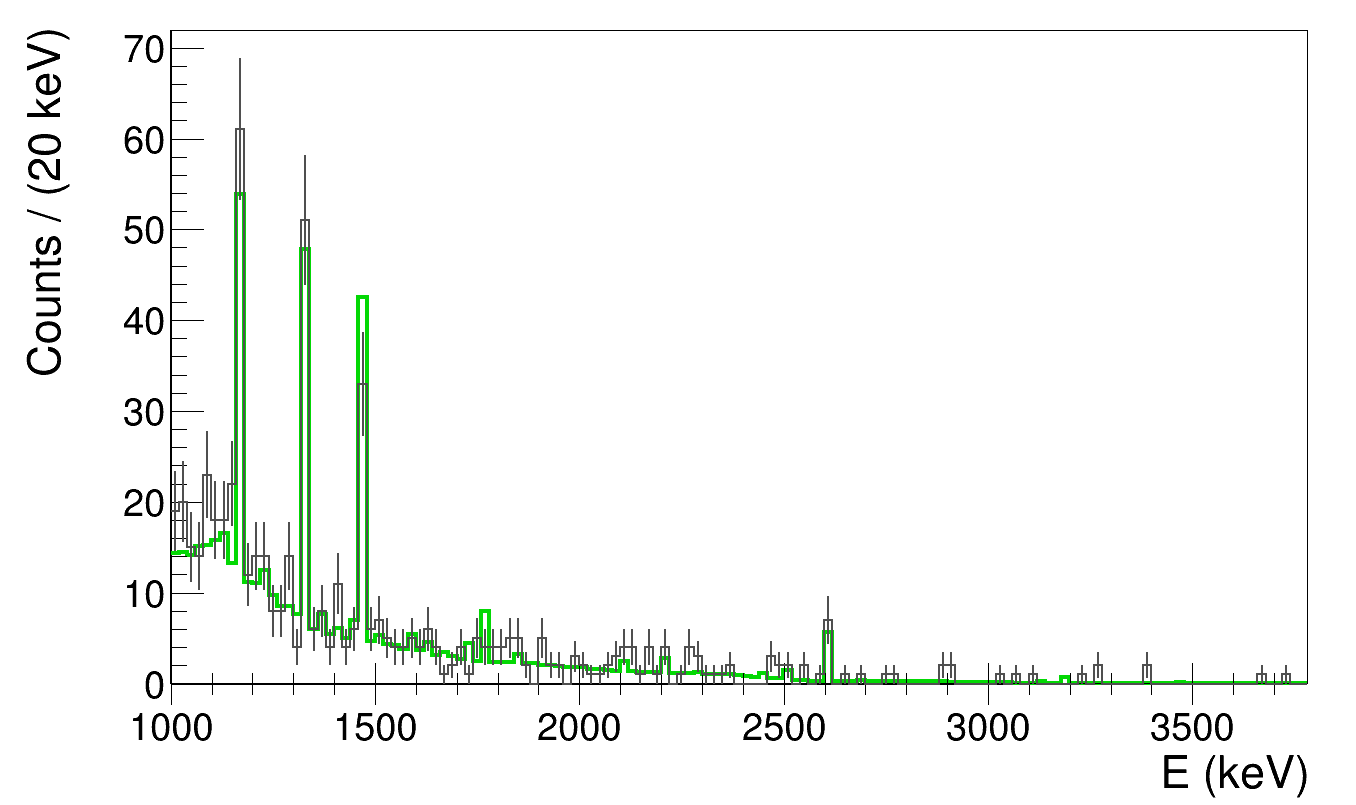}
\caption{\textbf{Comparison between the measured and the simulated background spectra.} 
The measured emission spectrum is shown in the region of interest as a dark grey histogram, with error bars representing one standard deviation. The simulated background distribution is shown in green for comparison. The simulation is based on a GEANT 4 validated MC characterization of the whole detector. The MC has as input the measured activities of the residual radionuclides, for each material present in the experimental setup. The simulation accounts for the emission probabilities and the decay schemes, the photon propagation and interactions in the materials of the apparatus and the detection efficiencies (see Section Methods).}
\label{comparison}
\end{figure}

The experimental and the MC simulated spectra agree to 88$\%$ in the energy range $\Delta E = (1000 \div 3800)~\mbox{keV}$, whereas in the low energy region there are larger deviations. This is mostly due to the impossibility to perfectly account for the residual cosmic rays and the bremsstrahlung caused by ${}^{210}$Pb and its daughters in the massive lead shield. The energy range  falls within the interval previously discussed for the validity of the theoretical model. 
Therefore we take $\Delta E$ as the energy Region Of Interest (ROI) for the following statistical analysis, the ROI is represented by the grey area in Fig. \ref{measured1}. 
In Fig. \ref{comparison} the measured spectrum is compared, in the ROI, with the simulated background distribution. 
The total number of simulated background counts within $\Delta E$ is $z_b=506$ events, to be compared with the measured number $z_c=576$ events. The reason for this low rate consists in the fact that the detector setup is especially designed for ultra low background measurements. The spectrum in 
Fig. \ref{measured1} only contains “real” events as the digital DAQ system has a filter rejecting noise events, by their pulse shape, with efficiency better then 99\%.

Then, we estimated the number of signal events which would be measured during the acquisition time, generated in the materials of the apparatus as collapse-induced photons. To this end the detection efficiencies were taken into account, which are shown, for the setup components which give an appreciable contribution, in Fig.~1 of the Supplementary Information. 

Given the rate in Eq.~\eqref{eq:rate} the expected signal contribution $z_s(R_0)$,  which is a function of the parameter $R_0$, turns out to be:
\begin{equation}\label{eq:p1}
z_s(R_0) = \sum_i \int_{\Delta E} \left.  \frac{d\Gamma_{t}}{dE} \right|_i T \epsilon_i(E) dE = \frac{a}{R_0^3},
\end{equation}
where $T$ is the total acquisition time of the experiment, $\epsilon_i(E)$ is the energy dependent efficiency function for the $i$-th component of the setup and $a\sim 1.8 \, \times 10^{-29}$ m$^3$. By substituting the values $z_c$, $z_b$ and $z_s$ in the $pdf$ of the parameter $R_0$ the following constraint is obtained:
\begin{equation}\label{eq:lowbound}
R_0> 0.54 \times 10^{-10}\, \text{m}
\end{equation}  
with probability 0.95. The data analysis is extensively described in the Supplementary Information, where the {\it pdf} is explicitly derived. 

It is important to stress that the energy range in which spontaneous photon emission is expected, extends from the upper threshold of the detector sensitive region (3.8 MeV) to 100 MeV (according to the emission rate given in Eq. (4)). A fraction of these \emph{primary} photons could be degraded in energy due to Compton scattering, thus producing  additional events in the ROI. Such a process would result in a stronger lower bound on $R_0$. We made an estimate of the improvement ($I$) on the bound  by considering the limiting case in which \emph{all} the primary spontaneously emitted photons generated in the $i$-th component of the setup, in the energy range $(3.8 \div 100) $ MeV,
are degraded, due to scattering, to the energy $E_i^{max, \,eff}$ within the ROI, corresponding to the maximal detection efficiency for the $i$-th material. 
We obtain $I \sim 1.620$, which is not sizeable (even under the exaggerated assumptions we considered); this is mainly due to the fact that spontaneous emission decreases with energy as $1/E$.

\begin{figure}[H]
\centering
\includegraphics[width=10cm]{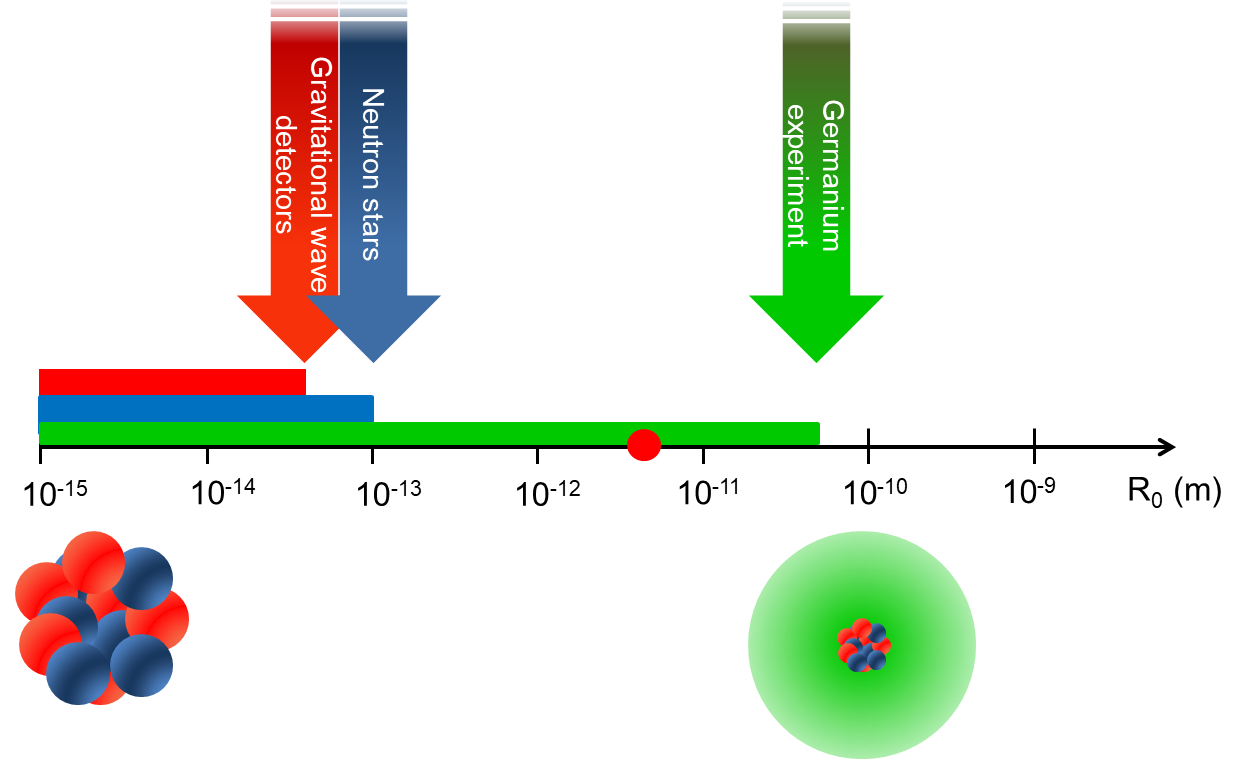}
\caption{\textbf{Lower bounds on the spatial cut-off $R_0$ of the Di\'osi-Penrose (DP) model.} According to Penrose, the value of $R_0$ is estimated to be the size of the wave function of the nuclei of the system, which amounts to $R_0 = 0.05 \times 10^{-10}$ m for the Ge crystal used in the experiment (red bullet on the horizontal scale). Our experiment sets a lower bound on $R_0$ at $0.54 \times 10^{-10}$ m (green bar and arrow), which is one order of magnitude larger than what predicted following Penrose's argument. Therefore, this parameter-free version of the DP model is excluded. The figure shows also previous lower bounds in the literature, similarly based on the monitoring of the Brownian-like diffusion predicted by the DP model. They refer to data analysis from gravitational wave detectors~\cite{helou2017lisa} ($R_0 \geq (40.1\pm 0.5) \times 10^{-15}$ m, red bar and arrow) and neutron stars~\cite{tilloy2019neutron} ($R_0 \gtrsim 10^{-13}$ m, blue bar and arrow).}
\label{boundR0}
\end{figure}

Our experiment sets a lower bound on $R_0$ of the order of $1\AA$, which is about three orders of magnitude stronger than previous bounds in the literature~\cite{helou2017lisa}; see Fig.~\ref{boundR0}. If $R_0$ is the size of the nucleus's wave function as suggested by Penrose, we have to confront our result with known properties of nuclei in matter. 
In a crystal, $R_0=\sqrt{\langle u^2 \rangle}$ where $\langle u^2 \rangle$ is the mean square displacement of a nucleus in the lattice, which can be computed by using the relation~\cite{debye1914verh, waller1923frage} $\langle u^2 \rangle=B/8\pi^2$, where $B = 0.20$ \AA$^2$ is the Debye-Waller factor for the Germanium crystal~\cite{gao1999parameterization}, cooled down at the liquid Nitrogen temperature. One obtains $R_0 = 0.05 \times 10^{-10}$ m, which is
more than one order of magnitude smaller than the lower limit set by our experiment. Therefore,
we conclude that Penrose's proposal for a gravity-related collapse of the wave function, in the
present formulation, is ruled out.

Of course, alternatives are always possible. Following Di\'osi, one option is to let $R_0$ completely free; but this comes at the price of having a parameter whose value is unjustified, apparently disconnected from the mass density of the system as well as from gravitational effects. Another option is to change the way the collapse is modeled (Poissonian decay), therefore adding extra terms and parameters to take into account a more complex dynamics, as done for other collapse models \cite{adler2007collapse, adler2008collapse, gasbarri2017gravity}. This kind of extensions was not envisaged in the literature so far.
Our result indicates that the idea of gravity-related wave function collapse, which remains very appealing, will probably require a radically new approach.


\section*{Acknowledgments}
The authors thank Prof. S.L. Adler, Prof. M. Arndt and Prof. H. Ulbricht for useful discussions and comments, and C. Capoccia, Dr. M. Carlesso and Dr. R. Del Grande for their help in preparing the figures. S.D. acknowledges  support from The Foundation BLANCEFLOR Boncompagni Ludovisi, n\'ee Bildt, INFN and the Fetzer Franklin Fund. K.P. and C.C. acknowledge the support of the Centro Fermi - Museo Storico della Fisica e Centro Studi e Ricerche ``Enrico Fermi'' (\emph{Open Problems in Quantum Mechanics project}), the John Templeton Foundation (ID 58158) and FQXi. L.D. acknowledges support of National Research Development and Innovation Office of Hungary Nos. 2017-1.2.1-NKP-2017-00001 and K12435, and support by FQXI minigrant. A.B. acknowledges support from the H2020 FET TEQ (grant n. 766900), the University of Trieste and INFN. All authors acknowledge support from the COST Action QTSpace (n. CA15220). 

\section*{Authors contributions} 
S.D. and A.B. conceived and designed the theoretical aspects of the research; M.L., K.P. and C.C. designed the experimental part of the research; S.D. performed the theoretical calculations, with the assistance of A.B. and L.D.; M.L., K.P. and C.C. performed the
experimental measurements; K.P. performed the data analysis, with the assistance of C.C., M.L. and L.D.; S.D., K.P. and A.B. prepared the manuscript and Supplementary Information in coordination with all authors.

\section*{\Large Methods}

\subsection{Calculation of the radiation emission rate.}

We summarize the main steps for deriving Eq.~(4)  for the emission rate of the main text.
The starting point is the quantum mechanical formula for the radiation emission rate
\begin{equation}
\frac{d}{d\omega}\Gamma_{t}=\frac{k^{2}}{c}\sum_{\mu}\int d\Omega_{k}\frac{d}{dt}\langle a_{\mathbf{k}\mu}^{\dagger}a_{\mathbf{k}\mu}\rangle_{t},\label{ratedef_ad}
\end{equation}
where $\langle a_{\mathbf{k}\mu}^{\dagger}a_{\mathbf{k}\mu}\rangle_{t}$
gives the average number of photons emitted at time $t$ with wave
vector $\boldsymbol{k}$ and polarization $\mu$. The time derivative
accounts for the fact that we are computing a rate; the
integration over the directions and polarizations of the photons for
the fact that we are interested in the total number of photons emitted
in a given energy range, independently from these degrees of freedom;
and the factor $\frac{k^{2}}{c}$ for the density of wave vectors
with modulus $k$. 

The expectation value $\langle a_{\mathbf{k}\mu}^{\dagger}a_{\mathbf{k}\mu}\rangle_{t}$ is computed starting from the master equation~(3) of the main text, which is convenient to rewrite in the following form~\cite{bahrami2014role}:
\begin{equation}
\frac{d\rho(t)}{dt}=-\frac{i}{\hbar}\left[H,\rho(t)\right]+\int d\boldsymbol{Q}\sum_{n,n'}\tilde{\Gamma}_{n,n'}(\boldsymbol{Q})\left(e^{\frac{i}{\hbar}\boldsymbol{Q}\cdot\boldsymbol{x}_{n}}\rho(t)e^{-\frac{i}{\hbar}\boldsymbol{Q}\cdot\boldsymbol{x}_{n'}}-\frac{1}{2}\left\{ e^{-\frac{i}{\hbar}\boldsymbol{Q}\cdot\boldsymbol{x}_{n'}}e^{\frac{i}{\hbar}\boldsymbol{Q}\cdot\boldsymbol{x}_{n}},\rho(t)\right\} \right)\label{ME_ad}
\end{equation}
where
\begin{equation}
\tilde{\Gamma}_{n,n'}(\boldsymbol{Q})=\frac{4G}{\pi\hbar^{2}}\frac{\tilde{\mu}_{n}(\boldsymbol{Q})\tilde{\mu}_{n'}^{*}(\boldsymbol{Q})}{Q^{2}}.
\end{equation}
with 
\begin{equation}
\tilde{\mu}(\boldsymbol{Q})=\frac{1}{2 \pi \hbar^3}\int d\boldsymbol{y}\mu(\boldsymbol{y})e^{-\frac{i}{\hbar}\boldsymbol{Q}\cdot\boldsymbol{y}}
\end{equation}
the Fourier transform of the mass density $\mu(\boldsymbol{y})$.

We moved to the Heisenberg picture, introducing the adjoint
master equation of Eq.~(\ref{ME_ad}), which for a generic operator $O$ takes the
form~\cite{Breuer:2002aa}: 
\begin{equation}
\frac{d}{dt}O(t)=\frac{i}{\hbar}\left[H,O(t)\right]+\int d\boldsymbol{Q}\sum_{k,k'}\tilde{\Gamma}_{k,k'}(\boldsymbol{Q})\left(e^{-\frac{i}{\hbar}\boldsymbol{Q}\cdot\boldsymbol{x}_{k'}}O(t)e^{\frac{i}{\hbar}\boldsymbol{Q}\cdot\boldsymbol{x}_{k}}-\frac{1}{2}\left\{ O(t),e^{-\frac{i}{\hbar}\boldsymbol{Q}\cdot\boldsymbol{x}_{k'}}e^{\frac{i}{\hbar}\boldsymbol{Q}\cdot\boldsymbol{x}_{k}}\right\} \right).\label{Osch_ad}
\end{equation}
The total Hamiltonian $H$ is the sum of three contributions:
\begin{equation}
H=H_{\text{\tiny S}}+H_{\text{\tiny R}}+H_{\text{\tiny INT}}.\label{Htot_ad}
\end{equation}
The first term is:
\begin{equation}
H_{\text{\tiny S}}=\sum_{j}\left(\frac{\boldsymbol{p}_{j}^{2}}{2m_{j}}+V(\boldsymbol{x}_{j})+\sum_{i<j}U(\boldsymbol{x}_{j}-\boldsymbol{x}_{i})\right)\label{Hsys_ad}
\end{equation}
where the sums run over all  particles of the system; $V$ is an
external potential and $U$ the potential among the particles of the system. We specialize on the emission from a crystal,
therefore the sum will run over all the electrons and the nuclei of the system 
(given the energy range of the emitted photons we consider, we do not need to resolve the internal structure of the nuclei by considering their protons). The free electromagnetic Hamiltonian is 
\begin{equation}
H_{\text{\tiny R}}=\sum_{\mu}\int d\boldsymbol{k}\hbar\omega_{k}\left(\frac{1}{2}+a_{\boldsymbol{k},\mu}^{\dagger}a_{\boldsymbol{k},\mu}\right)\label{Hem_ad}
\end{equation}
where $\omega_{k}=kc$ and $a_{\boldsymbol{k},\mu}$, $a_{\boldsymbol{k},\mu}^{\dagger}$
are, respectively, the annihilation and creation operators of a photon
with wave-vector $\boldsymbol{k}$ and polarization $\mu$. 
The last term describes the usual interaction between the electromagnetic field and the
particles (at the non-relativistic level): 
\begin{equation}
H_{\text{\tiny INT}}=\sum_{j}\left(-\frac{e_{j}}{m_{j}}\right)\boldsymbol{A}(\boldsymbol{x}_{j})\cdot\boldsymbol{p}_{j}+\sum_{j}\frac{e_{j}^{2}}{2m_{j}}\boldsymbol{A}^{2}(\boldsymbol{x}_{j}),\label{Hint_ad}
\end{equation}
where $e_{j}$, $m_{j}$ are the charge and the mass of the $j$-th
particle of the system and $\boldsymbol{A}(\boldsymbol{x})$ is the
vector potential which can be expanded in plane waves as:
\begin{equation}
\boldsymbol{A}(\boldsymbol{x})=\int d\boldsymbol{k}\sum_{\mu}\alpha_{k}\left[\vec{\epsilon}_{\boldsymbol{k},\mu}\,a_{\boldsymbol{k},\mu}\,e^{i\boldsymbol{k}\cdot\boldsymbol{x}}+\vec{\epsilon}_{\boldsymbol{k},\mu}\,a_{\boldsymbol{k},\mu}^{\dagger}\,e^{-i\boldsymbol{k}\cdot\boldsymbol{x}}\right]\label{A_ad}
\end{equation}
with $\alpha_{k}=\sqrt{\frac{\hbar}{2\varepsilon_{0}\omega_{k}(2\pi)^{3}}}$,
$\varepsilon_{0}$ the vacuum permittivity and $\vec{\epsilon}_{\boldsymbol{k},\mu}$
the (real) polarization vectors. Note that in Eq.~(\ref{Hint_ad}) the term
proportional to $\boldsymbol{p}_{j}\cdot\boldsymbol{A}(\boldsymbol{x}_{j})$
is missing because we are working in the Coulomb gauge where $\nabla\cdot\boldsymbol{A}=0$,
implying $\vec{\epsilon}_{\boldsymbol{k},\mu}\cdot\boldsymbol{k}=0$;
therefore this term contributes in the same way as the term $\boldsymbol{A}(\boldsymbol{x}_{j})\cdot\boldsymbol{p}_{j}$
and the two can be added.

Starting from Eq.~\eqref{Osch_ad}, the average number of photons emitted at time $t$ $\langle a_{\mathbf{k}\mu}^{\dagger}a_{\mathbf{k}\mu}\rangle_{t}$ is computed and then inserted in Eq.~(\ref{ratedef_ad}) to find the rate. 
The calculation is long but conceptually simple: the integral form of Eq.~\eqref{Osch_ad} is expanded perturbatively up to the second order (the first order terms give no contribution), similarly to what is usually done when the evolution operator is expanded using the Dyson series. Then one has to compute the nine terms resulting from this expansion. The calculation is fully reported in the Supplementary Information; here we give the physical picture underlying the calculations, which proved to be successful when applied to other models of spontaneous wave function collapse~\cite{fu1997spontaneous, rama, abd, bd, dirk}.

One can understand the mechanism of radiation emission in terms of a semi-classical picture. Each time there is a collapse, particles are ``kicked'', corresponding to an acceleration with associated radiation emission. 
The radiation emitted from different particles may add coherently or incoherently and to understand under which conditions they occur, it is instructive to study the radiation emission from two charged particles in the context of classical electrodynamics.

Suppose the two particles are accelerated by the same external force. At a point ``$x$'' very far away from the charges, the values of the emitted radiation crucially depends both on the distance $L$ between the particles  and the wavelength $\lambda$ of the emitted radiation.  
If the charges have opposite signs, when $L\ll \lambda$, given $x \gg L,\lambda$, the electric fields $E_1(x)$ generated by the positive charge and $E_2(x)$ generated by the negative charge will be the same, just with opposite sign, due to the opposite value of the charges. Then in this case $E_{\text{\tiny tot}}(x)= E_1(x)+ E_2(x)\simeq 0$, and because the emitted radiation is proportional to $|E_{\text{\tiny tot}}(x)|^2$, there is almost a full cancellation of the radiation field. 
On the contrary, if both charges have the same sign, $E_{\text{\tiny tot}}(x)= E_1(x)+ E_2(x)\simeq 2 E_1(x)$, hence the emitted radiation becomes four times larger than that emitted by a single charge. 
In more informal terms, we can say that for $L\ll \lambda$ a detector at $x$ sees the charges as if they are sitting in the same point. This leads to a coherent emission which suppresses the radiation emitted when the particles have opposite charges and maximizes it when they have the same charge. 

On the contrary, let us now consider the case $L\gg \lambda$. Still assuming $x \gg L,\lambda$, the two electric fields $E_1(x)$ and $E_2(x)$ have in general different intensities. In fact, if we label by $x_1 (x_2)$ the distance between the point ``$x$'' and the point where the first (second) particle is located, we have $|x_1-x_2|\sim L$. Then the electric fields oscillate many times in the distance $|x_1-x_2|$. Therefore, even if at a given point ``$x$'' they perfectly cancel, in a nearby point ``$x+dx$'' they add constructively. As for the intensity, one has $I(x) \propto |E_{\text{\tiny tot}}(x)|^2 = |E_1(x)|^2 + |E_2(x)|^2 + E^*_1(x)E_2(x)+ E_1(x)E^*_2(x)$, and when we integrate over a spherical surface of radius $|x|$, to find the total emission rate, the last two terms average to zero due to the fast oscillating behavior, and one gets that the two particles emit independently. 

Going back to the calculation in the main text, since the distance between electrons and nuclei is of order of one Angstrom, while the wavelength of the photons we are considering in the experiment is much smaller ($3.3 \times 10^{-3}<\lambda<1.2 \times 10^{-2}$ Angstrom), we are precisely in the second situations described here above, so electrons and nuclei emit independently. 
On the contrary, protons in the same nucleus are much closer than the smallest wavelength of the photons we are considering, which explains why they emit coherently.
As a result, the emission rate from the crystal is given by 
Eq.~(4) of the main text, where the emission from the electrons is neglected and the incoherent emission from all atoms in the crystal is considered.  

As a final note, in the DP model there is another reason for the incoherent radiation by electrons and nuclei, as long as $R_0 \ll L$ (which holds in our case). The gravitational fluctuations, underlying the decoherence term of Eq.~(3), which accelerate the charges, become uncorrelated beyond the range $R_0$: the electrons and the nuclei are accelerated by uncorrelated ``kicks'', resulting in an induced incoherent emission.

\subsection{Statistical analysis.}

Each component of the experimental apparatus was characterized by means of MC simulations (see \cite{boswell2010textsc})  based on the GEANT4 software library (verified by participating to international proficiency tests organised by the IAEA). 
The simulations were used to determine \emph{i}) the expected background due to residual radionuclides in the materials of the setup, \emph{ii}) the expected spontaneous radiation emission contribution to the measured spectrum.
More in detail:
\begin{itemize}

\item \emph{i}) the MC simulation of the background is based on the measured activities of the residual radionuclides, in all the components of the setup. The simulation accounts for the emission probabilities and the decay schemes, the photon propagation and interactions in the materials of the apparatus and the detection efficiencies. 
The obtained spectrum is compared with the measured distribution in Fig. \ref{comparison}.

\item \emph{ii}) the efficiency, as a function of the energy, for the detection of spontaneously emitted photons was obtained by generating 10$^8$ photons, for each component of the setup, in steps of 200
keV (i.e. 15 points in the ROI $\Delta E = E_1 \div E_2 = (1000 \div 3800)$ keV). 
The efficiency functions $\epsilon_i(E)$, $i$ labelling the material of the detector, were then estimated from polynomial fits of the corresponding distributions. Given the rate in Eq.~\eqref{eq:rate} one expects to measure a number of events:
\begin{equation}
\int_{\Delta E} \left.  \frac{d\Gamma_{t}}{dE} \right|_i T \epsilon_i(E) dE,
\end{equation}
due to the spontaneous emission by protons belonging to the $i$-th material, during the acquisition time $T$. Summing over all the materials, the total signal contribution (see Eq. \eqref{eq:p1}) is obtained: $z_s(R_0)=a/R_0^3$. 
\end{itemize}

The stochastic variable, representing the total number of photon counts measured in the range $\Delta E$, follows a Poisson distribution:
\begin{equation}\label{eq:zcdist}
p(z_{c}|\Lambda_c)=\frac{\Lambda_{c}^{z_{c}}e^{-\Lambda_{c}}}{z_{c}!},
\end{equation}
with $\Lambda_c$ the corresponding expected value.
Two sources contribute to the measured spectrum: a background ($b$) originated by all known emission processes, together with a potential signal ($s$) due to spontaneously emitted photons induced by the collapse process. The total number of counts, respectively $z_{b}$ and $z_{s}$, which would be measured in the period $T$, were estimated according to $i$ and $ii$. The corresponding independent stochastic variables can be also associated to Poisson distributions, whose expected values ($\Lambda_b$ and $\Lambda_s$) are then related by:
\begin{equation}\label{eq:expval}
\Lambda_c (R_0)=\Lambda_b+\Lambda_s(R_0)=z_b+z_s(R_0)+2
\end{equation}
where the dependence on $R_0$ is explicitly shown.

The $pdf$ of $\Lambda_c(R_0)$ can then be obtained from Eq.~\eqref{eq:zcdist} by applying the Bayes theorem:
\begin{equation}\label{eq:bayes}
\tilde{p}\left(\Lambda_c(R_0)|p(z_{c}|\Lambda_c(R_0))\right) = \frac{p(z_{c}|\Lambda_c(R_0)) \cdot \tilde{p}_0(\Lambda_c(R_0))}{\int_D p(z_{c}|\Lambda_c(R_0)) \cdot \tilde{p}_0(\Lambda_c(R_0)) \, d[\Lambda_c(R_0)]},
\end{equation} 
with $D$ the domain of $\Lambda_c$ and $\tilde{p}_0$ the prior distribution. 
$R_0$ is constrained by the requirement $R_0> R_0^{min} =10^{-14}$ m, which implies an upper bound on $\Lambda_c$ (see Eq. \eqref{eq:expval}). We then used a Heaviside function for the prior
\begin{equation}\label{eq:prior}
\tilde{p}_0(\Lambda_c(R_0))= \theta(\Lambda_c^{max}-\Lambda_c(R_0)),
\end{equation}
with $\Lambda_c^{max}=\Lambda_c(R_0^{min})$.
From Eq. \eqref{eq:bayes} the $pdf$ of $\Lambda_c(R_0)$ is:
\begin{equation}\label{eq:pdf}
\tilde{p}\left(\Lambda_c(R_0)\right) = \frac{\Lambda_{c}^{z_{c}} \, e^{-\Lambda_{c}} \, \theta(\Lambda_c^{max}-\Lambda_c)}{\int_0^{\Lambda_c^{max}} \Lambda_{c}^{z_{c}} \, e^{-\Lambda_{c}} \, d\Lambda_c}.
\end{equation} 
In order to obtain the bound given in Eq. \eqref{eq:lowbound} one then has to solve the following integral equation for the cumulative $pdf$: 
\begin{equation}\label{eq:cumulative}
\tilde{P}\left( \bar{\Lambda}_c \right) = \frac{\gamma(z_c+1,\bar{\Lambda}_c)}{\gamma(z_c+1,\Lambda_c^{max})}=0.95
\end{equation} 
which yields $\Lambda_{c} < \bar{\Lambda}_c = 617$. As a consequence
\begin{equation}\label{eq:limit}
\Lambda_c(R_0) = \Lambda_s(R_0) + \Lambda_b < 617 \Rightarrow \frac{a}{R_0^3} + \Lambda_b +1 < 617 \Rightarrow R_0 > \sqrt[3]{\frac{a}{616 - \Lambda_b}}.
\end{equation}

The analysis was performed in the energy range $E_1 \div E_2$ in which all the hypotheses of the model, for the spontaneous emission of protons, are fulfilled. However the energy range in which spontaneous photon emission is expected, according to Eq.~\eqref{eq:rate}, extends up to 100 MeV.
A fraction of spontaneously emitted photons with energy $E \in E_2 \div E_3 = (3.8 \div 100)$ MeV could be degraded in energy due to Compton scattering, thus contributing to $\Lambda_s(R_0)$; for this reason we estimated the corresponding improvement ($I$) to the bound  in Eq. \eqref{eq:lowbound}. 
Any improvement in the description of the expected background (or signal) contribution would lead to a bigger value of $\Lambda_b$ (or $a$), and from Eq. \eqref{eq:limit} one can infer that this would translate in a stronger bound on $R_0$.

Regarding $\Lambda_b$, since the MC simulation is based on the measured activities, we do not expect a contribution to photon emission, at energies higher than 3.8 MeV, originated from radionuclides decays.

The total number of spontaneously emitted photons which are generated in the materials of the detector in the energy range $E_2 \div E_3$ is given by:
\begin{equation}
\sum_i \int_{E_2}^{E_3} \left. \frac{d\Gamma_t}{dE} \right|_i T \, dE = 
 \sum_i \int_{E_2}^{E_3} N^2_{i} \, N_{ai} \, \beta \, T \, \frac{1}{R_0^3E}  dE
= \frac{b}{R_0^3} > 0,
\end{equation}
where $N_{i}$ and $N_{ai}$ are, respectively, the number of protons contained in each atom and the number of atoms of the $i$-th material, while the constant $\beta$ is defined as:
\begin{equation}
\beta = \frac{2}{3}\frac{G e^2}{\pi^{3/2} \varepsilon_0 c^3},
\end{equation}
$G, e, \varepsilon_0$ and $c$ are constants of nature with the usual meaning.
Let us indicate with $f$ the fraction of these photons which, due to Compton scattering, produce events in the ROI and are detected. The total signal contribution turns then to be:
\begin{equation}\label{3}
z_s(R_0) = \frac{a}{R_0^3} + \frac{f \, b}{R_0^3} >  \frac{a}{R_0^3}.
\end{equation}
Since $a + f \, b > a$, the contribution of the spontaneous emission in the range $E_2 \div E_3$ improves the bound on $R_0$ by a factor  $\sqrt[3]{(a + f\, b)/a}$.

We extracted the maximal improvement $I$ under the extreme - nontheless most conservative - assumption that \emph{all} the primary spontaneously emitted photons generated in the $i$-th material, in the energy range $E_2 \div E_3$,
are degraded, due to scattering, to the energy $E_i^{max, \,eff} \in E_1 \div E_2$ which corresponds to the maximal efficiency for the corresponding material (see Fig. 1 of the Supplementary Information). The total signal contribution then amounts to:
\begin{eqnarray}\begin{split}
z_s(R_0) = & \sum_i \int_{E_1}^{E_2} \left. \frac{d\Gamma}{dE} \right|_i T\, \epsilon_i (E) \, dE + \sum_i \epsilon_i^{max} \int_{E_2}^{E_3} \left. \frac{d\Gamma}{dE} \right|_i T\, dE  = \\ 
& = \left( (1.756 + 5.712)  \, \times 10^{-29} \right) \, \frac{\mathrm{m}^3}{R_0^3} = \frac{a + f \, b}{R_0^3}
\end{split}\end{eqnarray}
which corresponds to an improvement
$I \sim 1.620$. The improvement is not sizeable, as stated in the main text, even under the exaggerated assumptions we considered.

\newpage

\section*{\Large Supplementary Information}

\section{Statistical Analysis}

\subsection{Expected value of the integral number of detected photons\\ \\}\label{prob}

The measured emission spectrum, corresponding to a data taking period of about 62 days, is shown in Fig. 3.
As described in section \ref{bkg} a MC investigation of the background caused by known emission processes was performed. 88$\%$ of the measured counts can be interpreted in terms of background in the energy range $\Delta E = (1000 \div 3800)\; \mbox{keV}$. $\Delta E$ satisfies all the theoretical constraints on the spontaneous emission rate (Eq. (4) in the main text) provided that the dominant contribution of protons is considered only, since electrons are relativistic in this range. We than took $\Delta E$ as the region of interest for the following analysis. 

Following the notation introduced in the main text, we call $z_c$ the total number of experimentally measured photon counts in the energy range $\Delta E$, $z_c = 576$. The stochastic variable representing the measured number of photons follows a Poisson distribution:
\begin{equation}\label{3}
p(z_{c}|\Lambda_c)=\frac{\Lambda_{c}^{z_{c}}e^{-\Lambda_{c}}}{z_{c}!},
\end{equation}
where we represent the parameter of a Poisson distribution with the capital letter ($\Lambda$); $\Lambda_c$ is the expected value for the total number of measured counts in $\Delta E$. 

Two sources can be considered to contribute to the measured spectrum: a background ($b$) component accounting for all the known emission processes, together with a potential signal ($s$) of spontaneously emitted photons originated by the collapse process.
Since both the spontaneous and background radiations can be associated to Poissonian distributions we can rewrite the expected number of measured counts as   $\Lambda_c=\Lambda_b+\Lambda_s$. The expected number of signal counts $\Lambda_s$
can be predicted on the base of the collapse model, as a consequence $\Lambda_s$ and $\Lambda_c$ are functions of the model parameters; in the present analysis $\Lambda_s = \Lambda_s(R_0)$ and $\Lambda_c = \Lambda_c(R_0)$. 

In the following sections \ref{sig} and \ref{bkg} $\Lambda_s(R_0)$ and $\Lambda_b$ are estimated.

\subsubsection{Estimate of the signal contribution to the measured spectrum}\label{sig}

\begin{figure*}[htb]
\centering
\includegraphics[width=15cm]{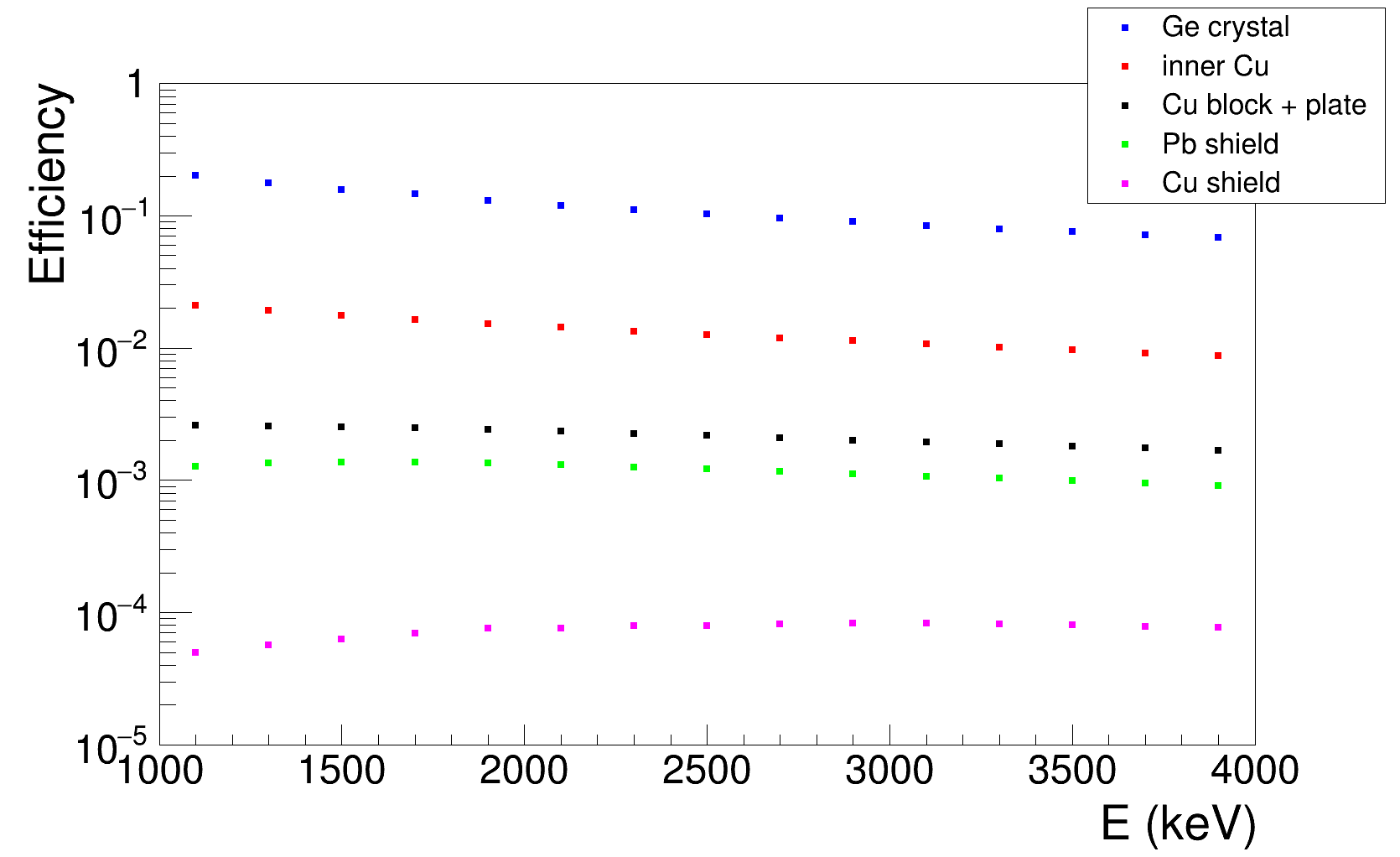}
\caption{\textbf{MC simulations of the photon detection efficiencies.}
The figure shows the detection efficiency distributions as a function of the energy in the ROI. The efficiencies were obtained by means of MC simulations, generating $10^8$ photons in each component of the detector, spaced by 200 keV. Colours are used to distinguish different materials of the experimental apparatus, as explained in the legend top right in the figure.}
\label{eff_ge}
\end{figure*}

In this section an estimate of the signal component will be given, i.e. the total number of spontaneously emitted photons which would be  measured by the Germanium detector during the acquisition time $T$, as a consequence of the contribution of all the protons inside the experimental apparatus. The contribution of each material of the setup to the spontaneous emission rate (Eq. (4) in the main text), depends on its mass, atomic weight and density. The detection efficiency for the emitted photons strongly depends on the composition and the geometry of the setup. MC simulations (see \cite{boswell2010textsc}), based on the GEANT4 software library were performed by generating $10^8$ photons in each component of the detector, spaced by 200
keV (i.e. 15 points in the ROI $\Delta E = (1000 \div 3800)$ keV).
The efficiency spectra in the region of interest $\Delta E$, for all the materials which give a significant contribution, are shown in Fig. \ref{eff_ge}.

Polinomial fits were performed of each efficiency distribution in order to obtain the efficiency functions $\epsilon_i(E)$ (the index $i$ labels the materials of the apparatus):
\begin{equation}\label{8}
\epsilon_i(E)=\sum_{j=0}^{c_i} \xi_{ij}E^j,
\end{equation}
where $c_i$ represents the degree of the polinomial expansion for the efficiency function of the $i$-th component, $\xi_{ij}$ is the matrix of the coefficients.

The total number of signal counts expected to be emitted by the $i$-th component, in the energy range $\Delta E$, is calculated by integrating the theoretical rate (Eq. (4) in the main text) over $\Delta E$, weighted with the corresponding efficiency function and multiplying for the acquisition time $T$. The contributions are then summed to get the total number of detected spontaneously emitted photons:
\begin{equation}\begin{split}\label{9}
z_s(R_0) & = \sum_i \int_{E_1}^{E_2} \left. \frac{d\Gamma_t}{dE} \right|_i T\, \epsilon_i (E) \, dE = \\ 
& = \sum_i \int_{E_1}^{E_2} N^2_{i} \, N_{ai} \, \beta \, T \, \frac{1}{R_0^3E} \, \sum_{j=0}^{ci} \xi_{ij}E^j \, dE \\
& = 1.756 \, \times 10^{-29} \, \frac{\mathrm{m}^3}{R_0^3} = \frac{a}{R_0^3}.
\end{split}\end{equation}
$N_{i}$ and $N_{ai}$ are, respectively, the number of protons contained in each atom and the number of atoms of the $i$-th material while the constant $\beta$ is defined as:
\begin{equation}\label{11}
\beta = \frac{2}{3}\frac{G e^2}{\pi^{3/2} \varepsilon_0 c^3}
\end{equation}
where $G, e, \varepsilon_0$ and $c$ are constants of nature with the usual meaning. In Eq. (\ref{9}) the dependence of the
total number of detected spontaneously emitted photons on the parameter $R_0$ appears explicitely. Since $z_s$ follows a Poissonian distribution the corresponding expected value is $\Lambda_s = z_s +1$.

\subsubsection{Background contribution to the measured spectrum}\label{bkg}

In order to evaluate the background from known emission processes, the activities of the residual radionuclides in the components of the setup were measured. The background characterization was performed by means of MC simulations of the decays of each radionuclide contained in each material, taking into account the emission probabilities and the decay schemes, the photon propagation and interactions inside the materials of the detector (giving rise to the continuum part of the background spectrum) as well as the detection efficiencies.
In Fig. 4 of the main text the measured spectrum (dark grey histogram with error bars), is compared, in the ROI, with the simulated background distribution (green curve).

Given a total number $N_{ij}$ of generated MC events for the $i$-th material and the $j$-th radionuclide the corresponding number of background counts is:
\begin{equation}\label{14}
z_{b,ij}=\frac{m_i \, A_{ij}\,T \, N_{rec,ij}}{N_{ij}},
\end{equation}
where $m_i$ is the mass of the $i$-th component, $A_{ij}$s are the measured activities and $N_{rec,ij}$ is the number of detected photons. 
The experimental and the MC simulated spectra agree at 88$\%$ for energies greater then 1 MeV whereas the low energy region has some deviations. This is mostly due to the impossibility to perfectly account for the residual cosmic rays and the bremsstrahlung caused by ${}^{210}$Pb and its daughters in the massive lead shield.
The total number of background counts in $\Delta E$ is estimated to be:
\begin{equation}\label{15}
 z_b =\sum_{i,j} z_{b,ij}=506,
\end{equation}
$z_b$ is also Poissonian, consequently the corresponding expected value is $\Lambda_b = z_b +1$.

The expected value for the integral number of detected photons in $\Delta E$ is then :
\begin{equation}\label{16}
 \Lambda_c(R_0) = \Lambda_s + \Lambda_b = \frac{a}{R_0^3} + 508.
\end{equation}
In the next section the probability distribution function ($pdf$) of $\Lambda_c(R_0)$ will be obtained, from which the lower limit on $R_0$ is extracted.

\subsubsection{Lower limit on the $R_0$ parameter of the model}\label{dp}

The $pdf$ of $\Lambda_c(R_0)$, which we will simply denote as $\Lambda_c$ in this section, can be obtained from Eq.~\eqref{3} by applying the Bayes theorem:
\begin{equation}\label{17}
\tilde{p}\left(\Lambda_c|p(z_{c}|\Lambda_c)\right) = \frac{p(z_{c}|\Lambda_c) \cdot \tilde{p}_0(\Lambda_c)}{\int_D p(z_{c}|\Lambda_c) \cdot \tilde{p}_0(\Lambda_c) \, d\Lambda_c}.
\end{equation} 
If no information or constraint is given on $\Lambda_c$, then the prior $\tilde{p}_0(\Lambda_c)$ should be taken constant, and the posterior would be a gamma $pdf$ for $\Lambda_c$:
\begin{equation}\label{18}
\tilde{p}\left(\Lambda_c|p(z_{c}|\Lambda_c)\right) = \frac{\Lambda_{c}^{z_{c}} \, e^{-\Lambda_{c}}}{\Gamma(z_c+1)},
\end{equation} 
with $\Gamma$ being the Gamma function.
In order to estimate what is the lower limit on the parameter $R_0$ which is compatible with the measured number of photons $z_c$ in $\Delta E$ and the estimated background $z_b$, within a probability of 0.95, we should then solve the integral equation:
\begin{equation}\label{19}
\tilde{P}\left( \bar{\Lambda}_c \right) = \frac{\int_0^{\bar{\Lambda}_c}\Lambda_{c}^{z_{c}} \, e^{-\Lambda_{c}} d\Lambda_c}{\Gamma(z_c+1)}=\frac{\gamma(z_c+1,\bar{\Lambda}_c)}{\Gamma(z_c+1)}=0.95,
\end{equation} 
with $\gamma$ being the upper incomplete gamma function, which yields $\Lambda_{c} < \bar{\Lambda}_c = 617$. 

Since $R_0$ is constrained by the requirement $R_0> R_0^{min} =10^{-14}$ m the following prior is to be considered accordingly:
  \begin{equation}\label{20}
    \tilde{p}_0(\Lambda_c)= \theta(\Lambda_c^{max}-\Lambda_c),
  \end{equation}
where $\theta$ is the Heaviside function and $\Lambda_c^{max}$ is defined as:
\begin{equation}\label{21}
   \Lambda_c^{max} = \frac{a}{[R_0^{min}]^3}+508.
\end{equation}
As a consequence the $pdf$ for $\Lambda_c$ assumes the expression:
\begin{equation}\label{22}
\tilde{p}\left(\Lambda_c|p(z_{c}|\Lambda_c)\right) = \frac{\Lambda_{c}^{z_{c}} \, e^{-\Lambda_{c}} \, \theta(\Lambda_c^{max}-\Lambda_c)}{\int_0^{\Lambda_c^{max}} \Lambda_{c}^{z_{c}} \, e^{-\Lambda_{c}} \, d\Lambda_c},
\end{equation} 
the integral equation to be solved for the cumulative $pdf$ is then: 
\begin{equation}\label{23}
\tilde{P}\left( \bar{\Lambda}_c \right) = \frac{\gamma(z_c+1,\bar{\Lambda}_c)}{\gamma(z_c+1,\Lambda_c^{max})}=0.95
\end{equation} 
The value of $\bar{\Lambda}_c$ is not significantly affected by the cutoff $R_0^{min}=10^{-14}$ m, introduced to guarantee that protons in the nuclei emit coherently. From the relation $\Lambda_{c}(R_0) < 617$ we then obtain: 
\begin{equation}\label{2}
\Lambda_s(R_0) + \Lambda_b < 617 \Rightarrow \frac{a}{R_0^3} + \Lambda_b +1 < 617 \;\;\Rightarrow\;\; R_0 > \sqrt[3]{\frac{a}{616 - \Lambda_b}}.
\end{equation}
which implies
\begin{equation}\label{24}
  R_0 > 0.54 \times 10^{-10}\; \textrm{m}
\end{equation}
with a probability of 0.95.

\clearpage

\section{Derivation of $\Delta E_{\text{\tiny DP}}$}
\label{sec:pen}

In deriving Eq.~(2) of the main text, Penrose assumes that the discrepancy between the spacetimes generated by two terms of a spatial superposition can be quantified, in the non-relativistic and Newtonian limit, by the expression
\begin{equation}
\Delta E_{\text{\tiny DP}}=\frac{1}{G}\int d\boldsymbol{r}\,\left(g_{a}(\boldsymbol{r})-g_{b}(\boldsymbol{r})\right)^{2},
\end{equation}
where $g_{a}(\boldsymbol{r})$ and $g_{b}(\boldsymbol{r})$ represent the accelerations experienced by a test mass at point $\boldsymbol{r}$, when the mass density of the system generating the gravitational field
is centered around  point $\boldsymbol{a}$ and $\boldsymbol{b}$ respectively.
The key idea is that, at $\boldsymbol{r}$, the square of the difference of the accelerations corresponding to each branch of the superposition is a good measure of how much the two space-times differ in that point. Then, the total difference is given by  integrating this quantity over space. The factor $1/G$, not present in the 1996 paper~\cite{penrose1996gravity} but later introduced~\cite{penrose2000wavefunction}, is required for dimensional reasons. 

Next, by using the relation $g(\boldsymbol{r})=-\nabla\Phi(\boldsymbol{r})$, performing an integration by parts and using the Poisson equation $\nabla^{2}\Phi(\boldsymbol{r})=4\pi G\mu(\boldsymbol{r})$
and its solution $\Phi(\boldsymbol{r})=-G\int d\boldsymbol{y}\frac{\mu(\boldsymbol{y})}{|\boldsymbol{r}-\boldsymbol{y}|}$, one arrives at the following relation:
\begin{equation}
\Delta E_{\text{\tiny DP}}=4\pi G\int d\boldsymbol{r}\int d\boldsymbol{r}'\frac{\left[\mu_{a}(\boldsymbol{r})-\mu_{b}(\boldsymbol{r})\right]\left[\mu_{a}(\boldsymbol{r}')-\mu_{b}(\boldsymbol{r}')\right]}{|\boldsymbol{r}-\boldsymbol{r}'|}.
\label{eq:De Pen}
\end{equation}
Here, $\mu_{a}$ and $\mu_{b}$ represent the mass density distribution of the system centered respectively around  position $\boldsymbol{a}$ and  position $\boldsymbol{b}$ in space, corresponding to the two terms of the superposition.  Note that, compared to the result in~\cite{penrose1996gravity}, Eq.~(\ref{eq:De Pen}) has a positive sign in front of it, which is required to avoid negative decay times of the superposition (see Eq.~(1) of the main text). 

A specific dynamic leading to a decay time of the form as in Eq.~(\ref{eq:De Pen}) were already introduced by Di\'osi\footnote{The factor in front of the integrals in Eq.~(\ref{eq:De Pen}) changes in different articles of Penrose: in the original derivation~\cite{penrose1996gravity} it is equal to $4\pi G$ as in the derivation here, in~\cite{penrose2000wavefunction} is just equal to $G$ while in~\cite{howl2019exploring} is equal to $4\pi G \gamma$ with $\gamma$ a constant later set $\gamma=1/(8\pi)$. This last choice is equivalent to that of Di\'osi in~\cite{diosi1987universal}, where the factor is equal to $G/2$.} in the form of a unitary stochastic model~\cite{diosi1987universal} and of a collapse model~\cite{diosi1989models}.

It is also interesting to note that, in more recent papers~\cite{penrose2014gravitization, howl2019exploring}, Penrose derives Eq.~(\ref{eq:De Pen}) through requiring the validity of equivalence principle at the quantum level. 

To conclude, in our analysis we assume that  $\mu_{a}(\boldsymbol{r})$ and $\mu_{b}(\boldsymbol{r})$ represent the same mass distribution (but differently located):  $\mu_{a}(\boldsymbol{r})=\mu(\boldsymbol{r}-\boldsymbol{a})$ and $\mu_{b}(\boldsymbol{r})=\mu(\boldsymbol{r}-\boldsymbol{b})$. This condition is fulfilled by rigid bodies, the kind of systems we consider in this work.  In this case Eq.~(\ref{eq:De Pen}) simplifies to Eq.~(2) of the main text, with $\boldsymbol{d}:=\boldsymbol{a}-\boldsymbol{b}$.

\section{Decay time $\tau_{\text{\tiny DP}}$ for a crystal structure}
\label{sec:mar}

Following Penrose~\cite{penrose2014gravitization, marshall2003towards}, we consider a mono-atomic simple cubic crystal, with lattice constant $a$. The contribution from the electrons can be neglected for two reasons: first, their mass is negligible compared to that of the nucleons; second, their wave function and therefore their mass distribution, according to Penrose, is much more spread out in space, making the self-gravitational energy smaller. 

\begin{figure*}[htb]
\centerline{%
\includegraphics[width=13cm]{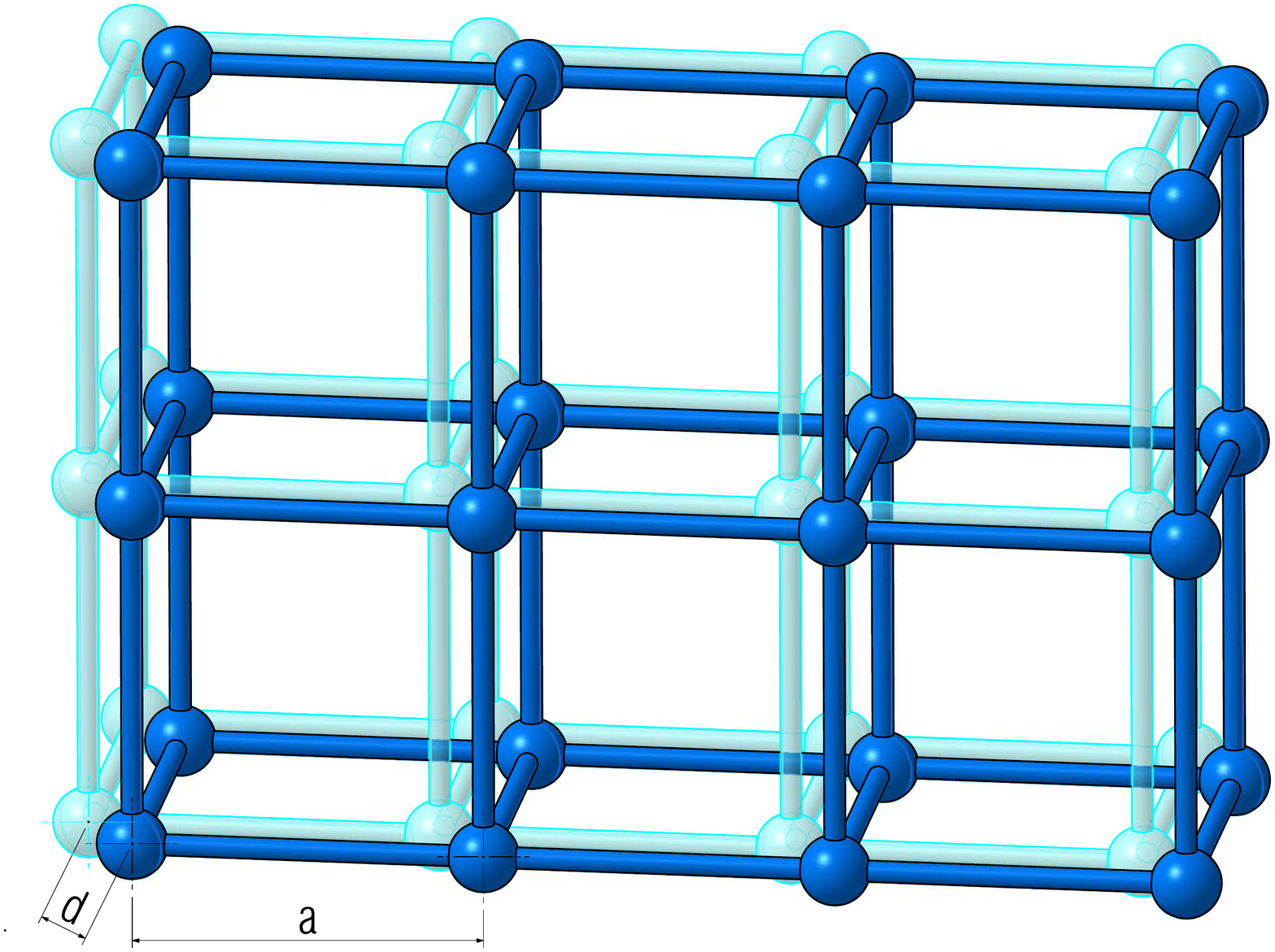}}
\caption{\textbf{Spatial superposition at distance $d$ of a crystal with lattice constant $a$.}
The figure shows a crystal lattice set in a spatial superposition at a distance $d$. A detailed proposal to achieve such a superposition for $d\simeq 10^{-13}$ m, using an optomechanical setup, was given in~\cite{marshall2003towards}.}

\label{crys_sup}
\end{figure*}

Let us suppose the crystal is initially in a superposition of two different positions in space, separated by a distance $\boldsymbol{d}$ such that  $d = |\boldsymbol{d}| \gg R_0$ (as an example, $d \sim 10^{-13}$ m is considered in~\cite{marshall2003towards}), such that the mass distributions in the two terms of the superposition are non-overlapping; see Fig. \ref{crys_sup}.
The mass density distribution of the crystal is the sum of the mass density distributions of each nucleus of its $N$ atoms (we assume that each nucleus has the same mass density distribution): 
\begin{equation}
\mu(\boldsymbol{r})=\sum_{i=1}^{N}\mu_{R_{0}}(\boldsymbol{r}-\boldsymbol{x}_{i}),
\end{equation}
where $\boldsymbol{x}_{i}$ is the equilibrium position of the center of mass of the $i$-th nucleus and $\mu_{R_{0}}(\boldsymbol{r})$ is the mass distribution of a single nucleus. We model it as a sphere of radius $R_0$:
\begin{equation}
\label{eq:dfghs}
\mu_{R_{0}}(\boldsymbol{r})=\frac{3m}{4\pi R_{0}^{3}}\theta(R_{0}-r), \qquad r=|\boldsymbol{r}|.
\end{equation}
We first compute $\Delta E_{\text{\tiny DP}}(\boldsymbol{d})$ for a single nucleus in a superposition state and then generalise the result to the whole crystal. 

From Eq.~(2) of the main text we have:
\begin{align}
\Delta E_{\text{\tiny DP}}(\boldsymbol{d})&=-8\pi G\int d\boldsymbol{r}'\left[\mu_{R_{0}}(\boldsymbol{r}'+\boldsymbol{d})-\mu_{R_{0}}(\boldsymbol{r}')\right]I_{R_{0}}(\boldsymbol{r}')\nonumber\\
&=-8\pi G\int d\boldsymbol{r}'\mu_{R_{0}}(\boldsymbol{r}')\left[I_{R_{0}}(\boldsymbol{r}'-\boldsymbol{d})-I_{R_{0}}(\boldsymbol{r}')\right],\label{eq:DE(d)}
\end{align}
where we introduced 
\begin{equation}\label{Ir}
I_{R_{0}}(\boldsymbol{r}'):=\int d\boldsymbol{r}\, \frac{\mu_{R_{0}}(\boldsymbol{r})}{|\boldsymbol{r}-\boldsymbol{r}'|}
=\begin{cases}
\displaystyle \frac{3m}{2R_{0}^{3}}\left(R_{0}^{2}-\frac{r'^{2}}{3}\right) & \text{if}\; r'\leq R_{0},\\
\displaystyle \frac{m}{r'} & \text{if}\; r'\geq R_{0};
\end{cases}
\end{equation}
in the last step, we used Eq.~\eqref{eq:dfghs} for $\mu_{R_{0}}(\boldsymbol{r})$. Taking into account that the mass density $\mu_{R_{0}}(\boldsymbol{r}')$
in Eq.~(\ref{eq:DE(d)}) is such that $r'\leq R_{0}$, one gets:
\begin{equation}
I_{R_{0}}(\boldsymbol{r}'-\boldsymbol{d})-I_{R_{0}}(\boldsymbol{r}')=\begin{cases}
\displaystyle \frac{m}{R_{0}^{3}}\left(\boldsymbol{r}'\cdot\boldsymbol{d}-\frac{d^{2}}{2}\right) & \text{if}\; |\boldsymbol{r}'-\boldsymbol{d}|\leq R_{0}, \\
\displaystyle \frac{m}{|\boldsymbol{r}'-\boldsymbol{d}|}-\frac{3m}{2R_{0}^{3}}\left(R_{0}^{2}-\frac{r'^{2}}{3}\right) & \text{if}\; |\boldsymbol{r}'-\boldsymbol{d}|\geq R_{0}.
\end{cases}
\end{equation}
There are two limiting cases of interest.  The first case occurs when $d\ll R_{0}$ and for almost\footnote{ ``Almost'' refers to the fact that the inequality is violated by some points for which $R_{0}\leq|\boldsymbol{r}'-\boldsymbol{d}|\leq R_{0}+d$. However, for small $d$, the contribution of these points is negligible.} all  points $\boldsymbol{r}'$ it holds $|\boldsymbol{r}'-\boldsymbol{d}|\leq R_{0}$. Then we have:
\begin{equation}\label{smalld}
\Delta E_{\text{\tiny DP}}(\boldsymbol{d})\simeq-\frac{8\pi Gm}{R_{0}^{3}}\int d\boldsymbol{r}'\mu_{R_{0}}(\boldsymbol{r}')\left(\boldsymbol{r}'\cdot\boldsymbol{d}-\frac{d^{2}}{2}\right)=\frac{4\pi Gm^{2}d^{2}}{R_{0}^{3}}.
\end{equation}
In the second case, the one we are interested in, $d\geq2R_{0}$, so that $|\boldsymbol{r}'-\boldsymbol{d}|\geq R_{0}$; one then has:
\begin{equation}
\Delta E_{\text{\tiny DP}}(\boldsymbol{d})=-8\pi G\int d\boldsymbol{r}'\mu_{R_{0}}(\boldsymbol{r}')\left[\frac{m}{|\boldsymbol{r}'-\boldsymbol{d}|}-\frac{3m}{2R_{0}^{3}}\left(R_{0}^{2}-\frac{r'^{2}}{3}\right)\right] = \frac{8\pi Gm^{2}}{R_{0}}\left(\frac{6}{5}-\frac{R_{0}}{d}\right).
\label{eq:DEspherefinal}
\end{equation}
Note that in the first case one has  a quadratic increase of the collapse time  with the square of the superposition distance, which is typical of decoherence phenomena in the regime of small spatial superpositions~\cite{schlosshauer2007decoherence}. In the second case, the collapse time saturates to a finite value; this is also a typical feature in open quantum system's dynamics e.g. collisional decoherence. Results in Eqs.~(\ref{smalld}), (\ref{eq:DEspherefinal}) are in agreement with those in~\cite{penrose2014gravitization, howl2019exploring}, apart for a factor $8 \pi$ due to a different definition of $\Delta E_{\text{\tiny DP}}$ used in these references, precisely by the same factor.

Considering the whole crystal we have:
\begin{equation}\label{DeDeDe}
\Delta E_{\text{\tiny DP}}(\boldsymbol{d})=-8\pi G\sum_{i,j=1}^{N}\int d\boldsymbol{r}\int d\boldsymbol{r}' \, \frac{\mu_{R_{0}}(\boldsymbol{r}-\boldsymbol{x}_{i})\left[\mu_{R_{0}}(\boldsymbol{r}'+\boldsymbol{d}-\boldsymbol{x}_{j})-\mu_{R_{0}}(\boldsymbol{r}'-\boldsymbol{x}_{j})\right]}{|\boldsymbol{r}-\boldsymbol{r}'|};
\end{equation}
we focus only on the case $d\geq2R_{0}$, which is consistent with the above mentioned values $d\sim10^{-13}$~m and the fact that the nucleus dimensions are of the order $R_0\sim10^{-15}-10^{-14}$~m. The contribution of all  terms with $i=j$ is given by Eq.~(\ref{eq:DEspherefinal}). The contribution of the terms with $i\neq j$ is negligible compared to that from the diagonal terms, as shown at the end of this section. Therefore we have:
\begin{equation}
\Delta E_{\text{\tiny DP}}(\boldsymbol{d})=N\frac{8\pi Gm^{2}}{R_{0}}\left(\frac{6}{5}-\frac{R_{0}}{d}\right).
\label{eq:jdfgjutk}
\end{equation}
Note that this formula cannot be obtained by simply replacing the single particle mass $m$ with the total mass $M=Nm$ of the system in Eq.~(\ref{eq:DEspherefinal}). In that case one would get a factor $N^2$ in place of $N$. This is due to the fact that $\Delta E_{\text{\tiny DP}}$ does not depend only on the mass of the object and the distance between the different branches of the superposition, but also on {\it how} the mass is spatially distributed. The reason why here we get a linear scaling in $N$ is related to the fact that $R_0\ll a$ and thus the contribution from the off diagonal terms of Eq.~(\ref{DeDeDe}) is negligible (see below). On the contrary, if $R_0\simeq a$, as for a fully homogeneous body, the contribution from the off diagonal terms would have been of the same order as that from the diagonal one, and we would have obtained a factor $N^2$ in Eq.~(\ref{eq:jdfgjutk}), making $\Delta E_{\text{\tiny DP}}$ proportional to the square of the total mass, consistently with Eq.~(\ref{eq:DEspherefinal}).

In the specific situation considered in~\cite{marshall2003towards}, the speck of matter which is in the superposition state at distance $d=10^{-13}$ m is a mirror with total mass  $M\simeq5\times10^{-12}$ Kg and the number of atoms is of order $N\simeq10^{14}$, implying that each nucleus has a mass $m\simeq5\times10^{-26}$ Kg. From this and Eq.~\eqref{eq:jdfgjutk}, choosing $R_0=10^{-14}$ m, one gets:
\begin{equation}\label{end_ad}
\Delta E_{\text{\tiny DP}}(\boldsymbol{d}) \simeq 4.61\times10^{-32}\,\textrm{J},
\end{equation}
corresponding to a decay time 
\begin{equation}
\tau_{\text{\tiny DP}}=\frac{\hbar}{\Delta E_{\text{\tiny DP}}(\boldsymbol{d})}\simeq0.0023 \,\,\,\textrm{s}.
\end{equation}
This decay time is about two orders of magnitudes smaller than the value usually considered by Penrose, which is $\tau_{\text{\tiny DP}}\sim 0.1$ s. If however, following~\cite{penrose2014gravitization}, the superposition is taken at a distance $d\simeq10^{-14}$ m, then the decay time becomes:
\begin{equation}
\tau_{\text{\tiny DP}}=\frac{\hbar}{\Delta E_{\text{\tiny DP}}(\boldsymbol{d})}\simeq0.013 \,\,\,\textrm{s}
\end{equation}
which is closer to the value suggested by Penrose.

\subsection{Contribution from the off diagonal terms in Eq.~(\ref{DeDeDe})}
We now set an upper bound on the contribution of the off diagonal terms in Eq.~(\ref{DeDeDe}) i.e.:
\begin{equation}\label{D_ad}
D:=-8\pi G I
\end{equation}
with
\begin{equation}
I:=\sum_{i\neq j}^{N}\int d\boldsymbol{y}\int d\boldsymbol{y}'\frac{\mu_{R_{0}}(\boldsymbol{y})[\mu_{R_{0}}(\boldsymbol{y}'+\boldsymbol{d})-\mu_{R_{0}}(\boldsymbol{y}')]}{|\boldsymbol{y}-\boldsymbol{y}'+\boldsymbol{x}_{i}-\boldsymbol{x}_{j}|}
\end{equation}
where, starting from Eq.~(\ref{DeDeDe}), we performed the change of variables $\boldsymbol{y}=\boldsymbol{r}-\boldsymbol{x}_i$ and $\boldsymbol{y}'=\boldsymbol{r}'-\boldsymbol{x}_j$. 

We introduce $\boldsymbol{x}_{ij}:=\boldsymbol{x}_{i}-\boldsymbol{x}_{j}$
and $x_{ij}=|\boldsymbol{x}_{ij}|$ and use the fact that $R_{0}\ll d\ll x_{ij}$
(in the first and third steps): 
\begin{align}
I&=\sum_{i\neq j}^{N}\int d\boldsymbol{y}\int d\boldsymbol{y}'\frac{\mu_{R_{0}}(\boldsymbol{y})[\mu_{R_{0}}(\boldsymbol{y}'+\boldsymbol{d})-\mu_{R_{0}}(\boldsymbol{y}')]}{|\boldsymbol{y}-\boldsymbol{y}'+\boldsymbol{x}_{i}-\boldsymbol{x}_{j}|}\simeq\sum_{i\neq j}^{N}\left(\frac{m^2}{|\boldsymbol{d}+\boldsymbol{x}_{ij}|}-\frac{m^2}{|\boldsymbol{x}_{ij}|}\right)\nonumber\\
&=\sum_{i\neq j}^{N}\frac{m^2}{x_{ij}}\left(\frac{1}{\sqrt{\left(\frac{d}{x_{ij}}\right)^{2}+1}}-1\right)\simeq\sum_{i\neq j}^{N}\frac{m^2}{x_{ij}}\left(-\frac{1}{2}\left(\frac{d}{x_{ij}}\right)^{2}\right)=-\frac{d^{2}m^2}{2}\sum_{i\neq j}^{N}\frac{1}{x_{ij}^{3}}.
\end{align}
Considering a cubic geometry of the crystal, we write: 
\[
\boldsymbol{x}_{i}=i_{1}a\hat{\boldsymbol{x}}+i_{2}a\hat{\boldsymbol{y}}+i_{3}a\hat{\boldsymbol{z}},
\]
with $i_{1},\,i_{2}$ and $i_{3}$ integers labelling the site where
the $i$-th nucleus is located. The conclusions of this analysis, are not affected by the specific choice of this geometry, what is relevant here are the differences in magnitude among $a$, $d$ and $R_{0}$. Then 
\[
x_{ij}=a\sqrt{(i_{1}-j_{1})^{2}+(i_{2}-j_{2})^{2}+(i_{3}-j_{3})^{2}},
\]
and 
\[
I\simeq-\frac{d^{2}m^2}{2a^{3}}S,
\]
with
\[
S:=\sum_{i=1}^{N}\sum_{j=1, (j\neq i)}^{N}\frac{1}{\left[(i_{1}-j_{1})^{2}+(i_{2}-j_{2})^{2}+(i_{3}-j_{3})^{2}\right]^{\frac{3}{2}}}.
\]

We now provide an upper bound for $S$. Because of the double sum, each nucleus interacts with
all the others. However, the terms of the sums between nuclei at
large distances are smaller than those between closer nuclei. This implies that the largest term of the sum over $i$ is
that given by the nucleus in the center of the cube. Therefore:
\begin{equation}
S\ll N\underset{(j_{1},j_{2},j_{3})\neq(0,0,0)}{\sum_{j_{1}=-n/2}^{n/2}\;\sum_{j_{2}=-n/2}^{n/2}\;\sum_{j_{3}=-n/2}^{n/2}}\frac{1}{\left(j_{1}^{2}+j_{2}^{2}+j_{3}^{2}\right)^{\frac{3}{2}}},\label{S_risp}
\end{equation}
where we introduced $n=\sqrt[3]{N}-1$, which is consistent with the
fact that the sum over $j$ runs over $N-1$ particles. Since the sums in Eq.~(\ref{S_risp}) are symmetric under the change $j_{\ell}\rightarrow-j_{\ell}$
(with $\ell=1,2,3$), one can also write 
\begin{align}\label{S2_risp}
S&\ll N\,2^{3}\underset{(j_{1},j_{2},j_{3})\neq(0,0,0)}{\sum_{j_{1}=0}^{n/2}\;\sum_{j_{2}=0}^{n/2}\;\sum_{j_{3}=0}^{n/2}}\frac{1}{\left(j_{1}^{2}+j_{2}^{2}+j_{3}^{2}\right)^{\frac{3}{2}}}\\
&=N 2^3 \left(\underset{(j_{1},j_{2},j_{3})\neq(0,0,0)}{\sum_{j_{1}=0}^{1}\;\sum_{j_{2}=0}^{1}\;\sum_{j_{3}=0}^{1}}+3\sum_{j_{1}=0}^{1}\sum_{j_{2}=0}^{1}\sum_{j_{3}=2}^{n/2}+3\sum_{j_{1}=0}^{1}\sum_{j_{2}=2}^{n/2}\sum_{j_{3}=2}^{n/2}+\sum_{j_{1}=2}^{n/2}\sum_{j_{2}=2}^{n/2}\sum_{j_{3}=2}^{n/2}\right)\frac{1}{\left(j_{1}^{2}+j_{2}^{2}+j_{3}^{2}\right)^{\frac{3}{2}}}\nonumber
\end{align}
The first term can be computed directly and it is equal to $3+\frac{3}{2\sqrt{2}}+\frac{1}{3\sqrt{3}}\simeq4.25$. For the other terms, one can make use of the fact that the function in the sums of $j_{1}$, $j_{2}$ and
$j_{3}$ is monotocally decreasing, and so for any $a\geq 0$ and $n>k>0$ 
\begin{equation}
\sum_{j=k}^{n}\frac{1}{\left(a+j^{2}\right)^{\frac{3}{2}}}\leq\int_{k-1}^{n}dx\frac{1}{\left(a+x^{2}\right)^{\frac{3}{2}}}.\label{Appro_risp}
\end{equation}
Using Eq.~(\ref{Appro_risp}) in each sum in Eq.~(\ref{S2_risp})
and changing to polar coordinates, we get
\begin{equation}
S\ll N\,2^{3}\left[4.25+3\int_{1}^{n/2}dx\left(\frac{1}{x^{3}}+\frac{2}{\left(1+x^{2}\right)^{\frac{3}{2}}}+\frac{1}{\left(2+x^{2}\right)^{\frac{3}{2}}}\right)+\right.\label{S3_risp}
\end{equation}
\[
\left.+3\frac{\pi}{2}\int_{1}^{n/\sqrt{2}}dr\left(\frac{1}{r^{2}}+\frac{r}{\left(1+r^{2}\right)^{\frac{3}{2}}}\right)+ \frac{\pi}{2}\int_{1}^{\sqrt{3}n/2}dr\frac{1}{r}\right]
\]
where for the last two terms we choose the range of radius of integration in such a way that the new integration region embeds the original one. We focus on each integral:
\[
3\int_{1}^{n/2}dx\left(\frac{1}{x^{3}}+\frac{2}{\left(1+x^{2}\right)^{\frac{3}{2}}}+\frac{1}{\left(2+x^{2}\right)^{\frac{3}{2}}}\right)=\frac{6n}{\sqrt{n^{2}+4}}+\frac{3n}{2\sqrt{n^{2}+8}}-\frac{6}{n^{2}}-3\sqrt{2}-\frac{\sqrt{3}}{2}+\frac{3}{2}\simeq
\]
\[
\simeq 9 -3\sqrt{2}-\frac{\sqrt{3}}{2}\simeq3.89\,,
\]
\[
3\frac{\pi}{2}\int_{1}^{n/\sqrt{2}}dr\left(\frac{1}{r^{2}}+\frac{r}{\left(1+r^{2}\right)^{\frac{3}{2}}}\right)=\frac{3}{4}\pi\left(-\frac{2\sqrt{2}}{\sqrt{n^{2}+2}}-\frac{2\sqrt{2}}{n}+\sqrt{2}+2\right)\simeq\frac{3}{4}\pi\left(\sqrt{2}+2\right)\simeq8.04\]
and
\[
\frac{\pi}{2}\int_{1}^{\sqrt{3}n/2}dr\frac{1}{r}=\frac{\pi}{2}\ln\left(\frac{\sqrt{3}n}{2}\right)
\]
Substituting in Eq.~(\ref{S3_risp})
\[
S\ll N\,2^{3}\left(17+\frac{\pi}{2}\ln[\frac{\sqrt{3}}{2}(\sqrt[3]{N}-1)]\right)\sim\frac{4\pi}{3}N\ln\left(N\right)\sim10^{16},
\]
for a crystal with $N=10^{14}$ atoms, which implies 
\[
|I|=\frac{d^{2}m^2}{2a^{3}}|S|\ll\frac{10^{-26} (5\times10^{-26})^2}{2\times10^{-30}}10^{16}\sim10^{-31}\; \textrm{Kg}^{2}\,\textrm{m}^{-1},
\]
which results in 
\begin{equation}
|D|=8\pi G |I|\ll 1.7 \times 10^{-40}\; \textrm{J},
\end{equation}
much smaller than the contributions from the terms with $i=j$, which is given in Eq.~(\ref{end_ad}).

\section{The Master Equation}\label{MEsupp}
In this section, we show how Eqs.~(1)  and~(2) necessarily imply Eq. (3) of the main text, if one phenomenologically assumes a Poissonian decay of superpositions as done in~\cite{howl2019exploring}. 

A  Poissonian collapse implies that in an infinitesimal time $dt$ there is
a probability $\lambda dt$ of having a collapse which will map $\rho(t)\rightarrow\mathcal{G}[\rho(t)]$,
with $\mathcal{G}$ the superoperator describing the effect of a collapse
on a given state $\rho(t)$; and a probability $(1-\lambda dt)$ of having no collapse, therefore the standard Schr\"odinger evolution. In mathematical terms:
\begin{equation}
\rho(t+dt)=(1-\lambda dt)\left[\rho(t)-\frac{i}{\hbar}\left[H,\rho(t)\right]dt\right]+\lambda dt\mathcal{G}[\rho(t)],\label{poisson}
\end{equation}
which can be easily rewritten as
\begin{equation}
\frac{d\rho(t)}{dt}=-\frac{i}{\hbar}\left[H,\rho(t)\right]+\lambda\left(\mathcal{G}[\rho(t)]-\rho(t)\right).\label{poisson2}
\end{equation}
The superoperator $\mathcal{L}[\rho(t)]:=\lambda\left(\mathcal{G}[\rho(t)]-\rho(t)\right)$ is known in the theory of open quantum systems as the ``Lindblad term". It must act linearly in $\rho(t)$ in
order to preserve the probabilistic interpretation of the statistical
operator, as well as to avoid faster than light signaling~\cite{gisin1989stochastic}. 

Since Eq.~(2) of the main text is translational invariant, a general theorem by Holevo~\cite{holevo1993note, holevo1993conservativity, vacchini2009quantum} fully characterizes the structure of Eq.~\eqref{poisson2}. The Lindblad term must be of the form: 
\begin{equation}
\mathcal{L}[\rho(t)]\!=\!\!\!\int\!\!\! d\boldsymbol{Q}\tilde{\Gamma}(\boldsymbol{Q})\sum_{j=1}^{\infty}\!\left[\!\left(e^{\frac{i}{\hbar}\boldsymbol{Q}\cdot\hat{\boldsymbol{x}}}L_{j}(\boldsymbol{Q},\hat{\boldsymbol{p}})\rho(t)L_{j}^{\dagger}(\boldsymbol{Q},\hat{\boldsymbol{p}})e^{-\frac{i}{\hbar}\boldsymbol{Q}\cdot\hat{\boldsymbol{x}}}-\frac{1}{2}\left\{ L_{j}^{\dagger}(\boldsymbol{Q},\hat{\boldsymbol{p}})L_{j}(\boldsymbol{Q},\hat{\boldsymbol{p}}),\rho(t)\right\}\! \right)\!\right]\!\!,\label{Holevo}
\end{equation}
with $\tilde{\Gamma}(\boldsymbol{Q})$ a function, $L_{j}(\boldsymbol{Q},\hat{\boldsymbol{p}})$ operators depending on the momentum $\hat{\boldsymbol{p}}$.

Gravity-related collapse makes no reference to the momentum
of the system, therefore $L_{j}(\boldsymbol{Q},\hat{\boldsymbol{p}})=L_{j}(\boldsymbol{Q})$. In this case $L_{j}(\boldsymbol{Q})$ become functions that can be reabsorbed
in the definition of $\tilde{\Gamma}(\boldsymbol{Q})$ (note that there are no problems with possible divergent quantities since, in writing  Eq.~(\ref{Holevo}), it is assumed that $\int d\boldsymbol{Q}\tilde{\Gamma}(\boldsymbol{Q})\sum_{j=1}^{\infty}|L_{j}(\boldsymbol{Q},\cdot)|^{2}<\infty$). Then, Eq.~(\ref{Holevo}) becomes:
\begin{equation}
\mathcal{L}[\rho(t)]=\int d\boldsymbol{Q}\tilde{\Gamma}(\boldsymbol{Q})\left(e^{\frac{i}{\hbar}\boldsymbol{Q}\cdot\hat{\boldsymbol{x}}}\rho(t)e^{-\frac{i}{\hbar}\boldsymbol{Q}\cdot\hat{\boldsymbol{x}}}-\rho(t)\right),\label{after Holevo}
\end{equation}
which implies that Eq.~(\ref{poisson2}) takes the form: 
\begin{equation}
\frac{d\rho(t)}{dt}=-\frac{i}{\hbar}\left[H,\rho(t)\right]+\int d\boldsymbol{Q}\tilde{\Gamma}(\boldsymbol{Q})\left(e^{\frac{i}{\hbar}\boldsymbol{Q}\cdot\hat{\boldsymbol{x}}}\rho(t)e^{-\frac{i}{\hbar}\boldsymbol{Q}\cdot\hat{\boldsymbol{x}}}-\rho(t)\right).
\end{equation}
In order to determine $\tilde{\Gamma}(\boldsymbol{Q})$ in agreement
with Penrose's decay rate (see Eqs.~(1) and (2) of the main text), we have to impose:
\begin{equation}
\langle\boldsymbol{a}|\mathcal{L}[\rho(t)]|\boldsymbol{b}\rangle=\left[\frac{8\pi G}{\hbar}\int d\boldsymbol{r}\int d\boldsymbol{r}'\frac{\mu(\boldsymbol{r}-\boldsymbol{a})\mu(\boldsymbol{r}'-\boldsymbol{b})-\mu(\boldsymbol{r})\mu(\boldsymbol{r}')}{|\boldsymbol{r}-\boldsymbol{r}'|}\right]\langle\boldsymbol{a}|\hat{\rho}_{t}|\boldsymbol{b}\rangle
\end{equation}
which, taking the matrix element of Eq.~(\ref{after Holevo}), implies
\begin{equation}
\frac{8\pi G}{\hbar}\int d\boldsymbol{r}\int d\boldsymbol{r}'\frac{\mu(\boldsymbol{r}-\boldsymbol{a})\mu(\boldsymbol{r}'-\boldsymbol{b})-\mu(\boldsymbol{r})\mu(\boldsymbol{r}')}{|\boldsymbol{r}-\boldsymbol{r}'|}=\int d\boldsymbol{Q}\tilde{\Gamma}(\boldsymbol{Q})\left(e^{\frac{i}{\hbar}\boldsymbol{Q}\cdot(\boldsymbol{a}-\boldsymbol{b})}-1\right).\label{sette}
\end{equation}
Introducing
\begin{equation}
\Gamma(\boldsymbol{y})=\frac{1}{(2\pi\hbar)^{3}}\int d\boldsymbol{Q}\tilde{\Gamma}(\boldsymbol{Q})e^{\frac{i}{\hbar}\boldsymbol{Q}\cdot\boldsymbol{y}},
\end{equation}
and $\boldsymbol{d}=\boldsymbol{a}-\boldsymbol{b}$, Eq.~(\ref{sette}) gives
\begin{equation}
\Gamma(\boldsymbol{d})=\frac{8\pi G}{\hbar(2\pi\hbar)^{3}}\int d\boldsymbol{r}\int d\boldsymbol{r}'\frac{\mu(\boldsymbol{r})\mu(\boldsymbol{r}'+\boldsymbol{d})}{|\boldsymbol{r}-\boldsymbol{r}'|}.
\end{equation}
Taking the Fourier transform of both sides and using the relation
\begin{equation}
\int d\boldsymbol{s}\frac{e^{-\frac{i}{\hbar}\boldsymbol{Q}\cdot\boldsymbol{s}}}{|\boldsymbol{s}|}=\frac{4\pi\hbar^{2}}{Q^{2}},
\end{equation}
it is straightforward to prove that 
\begin{equation}
\tilde{\Gamma}(\boldsymbol{Q})=\frac{8\pi G}{\hbar(2\pi\hbar)^{3}}\tilde{\mu}(\boldsymbol{Q})\frac{4\pi\hbar^{2}}{Q^{2}}\tilde{\mu}(-\boldsymbol{Q}).
\end{equation}
Simplifying the constants and using the property $\tilde{\mu}(-\boldsymbol{Q})=\tilde{\mu}(\boldsymbol{Q})^{*}$ (which follows from the fact that $\mu(\boldsymbol{r})$ is real) we get 
\begin{equation}
\tilde{\Gamma}(\boldsymbol{Q})=\frac{4G}{\pi\hbar^{2}}\frac{|\tilde{\mu}(\boldsymbol{Q})|^{2}}{Q^{2}}.\label{gammaqfinal}
\end{equation}
Note that compared to Eq.~(17) of~\cite{bahrami2014role} there
is a factor $8\pi$ difference: this is due to the fact that their
master equation differs from the one we are considering precisely by a factor $8\pi$
(compare also the Master Equation (6) of~\cite{bahrami2014role}
with the one here derived). 

To conclude, we proved that Eq.~(1) and~(2) of the main text, requiring only a Poissonian decay and translation
invariance, imply that the statistical operator for a single particle
follows the Master Equation (ME): 
\begin{equation}
\frac{d\rho(t)}{dt}=-\frac{i}{\hbar}\left[H,\rho(t)\right]+\int d\boldsymbol{Q}\tilde{\Gamma}(\boldsymbol{Q})\left(e^{\frac{i}{\hbar}\boldsymbol{Q}\cdot\hat{\boldsymbol{x}}}\rho(t)e^{-\frac{i}{\hbar}\boldsymbol{Q}\cdot\hat{\boldsymbol{x}}}-\rho(t)\right),
\end{equation}
with $\tilde{\Gamma}(\boldsymbol{Q})$ given in Eq.~(\ref{gammaqfinal}).
Performing the integration over $\boldsymbol{Q}$, one can rewrite this ME in the form:
\begin{equation}
\frac{d\rho(t)}{dt}=-\frac{i}{\hbar}\left[H,\rho(t)\right]-\frac{4\pi G}{\hbar}\int d\boldsymbol{x}\int d\boldsymbol{y}\frac{1}{|\boldsymbol{x}-\boldsymbol{y}|}\left[\mu(\boldsymbol{y},\hat{\boldsymbol{x}}),\left[\mu(\boldsymbol{x},\hat{\boldsymbol{x}}),\rho(t)\right]\right],
\end{equation} 
where $\mu(\boldsymbol{x},\hat{\boldsymbol{x}})$ represents the mass density distribution centered in $\hat{\boldsymbol{x}}$.

The generalization for multiparticle systems is straightforward, one just needs to replace the single particle mass density with the total mass density i.e. $\mu(\boldsymbol{x},\hat{\boldsymbol{x}}) \rightarrow \hat{M}(\boldsymbol{x})=\sum_{n}\mu_{n}(\boldsymbol{x},\hat{\boldsymbol{x}}_{n})$: 
\begin{equation}\label{S50}
\frac{d\rho(t)}{dt}=-\frac{i}{\hbar}\left[H,\rho(t)\right]-\frac{4\pi G}{\hbar}\int d\boldsymbol{x}\int d\boldsymbol{y}\frac{1}{|\boldsymbol{x}-\boldsymbol{y}|}\left[\hat{M}(\boldsymbol{y}),\left[\hat{M}(\boldsymbol{x}),\rho(t)\right]\right],
\end{equation}
which is precisely Eq.~(3) of the main text.

\section{Calculation of the radiation emission rate}

In this section Eq.~(4) of the main text, namely the power emission formula from a generic system, will be derived. 

The goal is to compute the radiation emission rate already introduced in the Methods in Eq. (7) and that we report here for convenience:
\begin{equation}
\frac{d}{d\omega_k}\Gamma_{t}=\frac{k^{2}}{c}\sum_{\nu}\int d\Omega_{k}\frac{d}{dt}\langle a_{\mathbf{k}\nu}^{\dagger}a_{\mathbf{k}\nu}\rangle_{t},\label{ratedef_ad}
\end{equation}
The starting point of our analysis is the adjoint master equation (ME) introduced in the methods in Eq. (11):
\begin{equation}
\frac{d}{dt}O(t)=\frac{i}{\hbar}\left[H,O(t)\right]+\int d\boldsymbol{Q}\sum_{k,k'}\tilde{\Gamma}_{k,k'}(\boldsymbol{Q})\left(e^{-\frac{i}{\hbar}\boldsymbol{Q}\cdot\boldsymbol{x}_{k'}}O(t)e^{\frac{i}{\hbar}\boldsymbol{Q}\cdot\boldsymbol{x}_{k}}-\frac{1}{2}\left\{ O(t),e^{-\frac{i}{\hbar}\boldsymbol{Q}\cdot\boldsymbol{x}_{k'}}e^{\frac{i}{\hbar}\boldsymbol{Q}\cdot\boldsymbol{x}_{k}}\right\} \right).\label{Osch_ad}
\end{equation}

\subsection{Perturbative Master Equation}

We start by moving to the interaction picture with respect to the system and
the free EM Hamiltonians, introducing $O^{I}(t)=e^{-\frac{i}{\hbar}(H_{\text{\tiny S}}+H_{\text{\tiny R}})t}Oe^{\frac{i}{\hbar}(H_{\text{\tiny S}}+H_{\text{\tiny R}})t}$,
so Eq.~(\ref{Osch_ad}) becomes:
\begin{equation}
\frac{d}{dt}O^{I}(t)=\mathcal{C}_{t}^{I}[O^{I}(t)]+\mathcal{L}_{t}^{I}[O^{I}(t)]\label{Oint_ad}
\end{equation}
where 
\begin{equation}
\mathcal{C}_{t}^{I}[O^{I}(t)]:=\frac{i}{\hbar}\left[H_{\text{\tiny INT}}^{I}(t),O^{I}(t)\right],\label{C_ad}
\end{equation}
\begin{equation}
\mathcal{L}_{t}^{I}[O^{I}(t)]:=\int d\boldsymbol{Q}\sum_{n,n'}\tilde{\Gamma}_{n,n'}(\boldsymbol{Q})\left(e^{-\frac{i}{\hbar}\boldsymbol{Q}\cdot\boldsymbol{x}_{n'}^{I}(t)}O^{I}(t)e^{\frac{i}{\hbar}\boldsymbol{Q}\cdot\boldsymbol{x}_{n}^{I}(t)}-\frac{1}{2}\left\{ O^{I}(t),e^{-\frac{i}{\hbar}\boldsymbol{Q}\cdot(\boldsymbol{x}_{n'}^{I}(t)-\boldsymbol{x}_{n}^{I}(t))}\right\} \right).\label{L_ad}
\end{equation}
The interaction picture introduced here has opposite signs in the exponents
compared to the standard one. This is a consequence of the fact that our
analysis is starting from the adjoint ME, which plays the role of the Heisenberg picture in the framework of open quantum systems.
In order to lighten the notation, the superscript ``$I$'' will be removed in what follows and the calculations are meant to be performed in the interaction picture. In addition, when the superoperators $\mathcal{C}_{t}$ and $\mathcal{L}_{t}$
apply to all the operators on their right, we will omit the square brackets denoting on which operators they are applied (e.g. $\mathcal{C}_{t_1}[\mathcal{L}_{t_2}[O(t)]]$ will be simply written as $\mathcal{C}_{t_1}\mathcal{L}_{t_2}O(t)$).

We start the standard perturbative expansion by integrating Eq.~(\ref{Oint_ad})
\begin{equation}
O(t)=O(0)+\int_{0}^{t}dt_{1}\mathcal{C}_{t_{1}}O(t_{1})+\int_{0}^{t}dt_{1}\mathcal{L}_{t_{1}}O(t_{1})\label{Oform_ad}
\end{equation}
and substituting this expression into each integral of Eq.~(\ref{Oform_ad})
\begin{equation}
O(t)=O(0)+\int_{0}^{t}dt_{1}\mathcal{C}_{t_{1}}O(0)+\int_{0}^{t}dt_{1}\mathcal{L}_{t_{1}}O(0)+\int_{0}^{t}dt_{1}\int_{0}^{t_{1}}dt_{2}\mathcal{C}_{t_{1}}\mathcal{C}_{t_{2}}O(t_{2})+\label{O2_ad}
\end{equation}
\[
+\int_{0}^{t}dt_{1}\int_{0}^{t_{1}}dt_{2}\mathcal{C}_{t_{1}}\mathcal{L}_{t_{2}}O(t_{2})+\int_{0}^{t}dt_{1}\int_{0}^{t_{1}}dt_{2}\mathcal{L}_{t_{1}}\mathcal{C}_{t_{2}}O(t_{2})+\int_{0}^{t}dt_{1}\int_{0}^{t_{1}}dt_{2}\mathcal{L}_{t_{1}}\mathcal{L}_{t_{2}}O(t_{2}).
\]
The last term can be neglected since we are interested to the lowest
relevant order in $G$. Substituting one more time, keeping only the
contributions of order $e^{2}$ and $G$, setting $t_{3}=0$ and $O(0)=O$
we get
\begin{align}
O(t)&=O+\int_{0}^{t}dt_{1}\mathcal{C}_{t_{1}}O+\int_{0}^{t}dt_{1}\mathcal{L}_{t_{1}}O+\int_{0}^{t}dt_{1}\int_{0}^{t_{1}}dt_{2}\mathcal{C}_{t_{1}}\mathcal{C}_{t_{2}}O+\int_{0}^{t}dt_{1}\int_{0}^{t_{1}}dt_{2}\mathcal{C}_{t_{1}}\mathcal{L}_{t_{2}}O\nonumber\\
&+\int_{0}^{t}dt_{1}\int_{0}^{t_{1}}dt_{2}\mathcal{L}_{t_{1}}\mathcal{C}_{t_{2}}O+\int_{0}^{t}dt_{1}\int_{0}^{t_{1}}dt_{2}\int_{0}^{t_{2}}dt_{3}\mathcal{C}_{t_{1}}\mathcal{C}_{t_{2}}\mathcal{L}_{t_{3}}O\nonumber\\
&+\int_{0}^{t}dt_{1}\int_{0}^{t_{1}}dt_{2}\int_{0}^{t_{2}}dt_{3}\mathcal{C}_{t_{1}}\mathcal{L}_{t_{2}}\mathcal{C}_{t_{3}}O+\int_{0}^{t}dt_{1}\int_{0}^{t_{1}}dt_{2}\int_{0}^{t_{2}}dt_{3}\mathcal{L}_{t_{1}}\mathcal{C}_{t_{2}}\mathcal{C}_{t_{3}}O\label{O3_ad}
\end{align}
which is the lowest order relevant contribution for the calculation of $O(t)$.

\subsection{Computation of the emission rate}

Let us consider the average number of emitted photons:
\begin{equation}
\langle a_{\mathbf{k}\nu}^{\dagger}a_{\mathbf{k}\nu}\rangle_{t}=\textrm{Tr}\left[\left(a_{\mathbf{k}\nu}^{\dagger}a_{\mathbf{k}\nu}\right)(t)\rho\right]\label{aat_ad}
\end{equation}
Note that since we are working with the adjoint equation, in principle
$\left(a_{\mathbf{k}\nu}^{\dagger}a_{\mathbf{k}\nu}\right)(t)$ should
represent the operators evolution in Heisenberg picture i.e. given
by Eq.~(\ref{Osch_ad}). However, moving to the interaction picture does
not change $\left(a_{\mathbf{k}\nu}^{\dagger}a_{\mathbf{k}\nu}\right)(t)$,
which is why we can proceed working in this picture. We assume $\rho_{i}=\rho_{g}\otimes|0\rangle\langle0|$
with $\rho_{g}$ the ground state of the system and $|0\rangle\langle0|$
the EM vacuum. 

Using Eq.~(\ref{O3_ad}) for $O=a_{\mathbf{k}\nu}^{\dagger}a_{\mathbf{k}\nu}$,
multiplying by $\rho_{i}$ and taking the trace, we can notice the following: $(i)$ the
first two terms are zero because of the average over the EM vacuum; $(ii)$ the third term is zero because $\mathcal{L}_{t_{1}}a_{\mathbf{k}\nu}^{\dagger}a_{\mathbf{k}\nu}=0$, since $a_{\mathbf{k}\nu}^{\dagger}a_{\mathbf{k}\nu}$ commute with
the system operators (so the first term in the round bracket of Eq.~(\ref{L_ad}) cancels with those coming from the anticommutator); $(iii)$ precisely for the same reason, also the fifth and the seventh terms
are zero; $(iv)$ the fourth term describes emission \textit{not} due to the
noise. Assuming, as we are doing, that the crystal in its ground state, makes this contribution equal to zero. If this is not the case, this term adds a contribution to the final emission rate formula Eq.~(4) in the main text, leading to a stronger bound on $R_{0}$. The assumption that the system is in the ground state is the most conservative one. 

Thus, we are left with 
\begin{equation}
\langle a_{\mathbf{k}\mu}^{\dagger}a_{\mathbf{k}\mu}\rangle_{t}=A_{1}+A_{2}+A_{3}\label{aat2_ad}
\end{equation}
where
\begin{align}
A_{1}&:=\int_{0}^{t}dt_{1}\int_{0}^{t_{1}}dt_{2}\textrm{Tr}\left[\rho_{i}\mathcal{L}_{t_{1}}\mathcal{C}_{t_{2}}a_{\mathbf{k}\nu}^{\dagger}a_{\mathbf{k}\nu}\right],\label{A1_ad}\\
A_{2}&:=\int_{0}^{t}dt_{1}\int_{0}^{t_{1}}dt_{2}\int_{0}^{t_{2}}dt_{3}\textrm{Tr}\left[\rho_{i}\mathcal{C}_{t_{1}}\mathcal{L}_{t_{2}}\mathcal{C}_{t_{3}}a_{\mathbf{k}\nu}^{\dagger}a_{\mathbf{k}\nu}\right],\label{A2_ad}\\
A_{3}&:=\int_{0}^{t}dt_{1}\int_{0}^{t_{1}}dt_{2}\int_{0}^{t_{2}}dt_{3}\textrm{Tr}\left[\rho_{i}\mathcal{L}_{t_{1}}\mathcal{C}_{t_{2}}\mathcal{C}_{t_{3}}a_{\mathbf{k}\nu}^{\dagger}a_{\mathbf{k}\nu}\right].\label{A3_ad}
\end{align}
The evaluation of these terms is given below, where we will show
that $A_{1}$ and $A_{2}$ give no contribution to the emission rate,
while the time derivative of $A_{3}$ is given by:
\begin{equation}
\frac{d}{dt}A_{3}=\frac{1}{\hbar^{2}\omega_{k}^{2}}\frac{4\pi G}{\hbar}\left(\frac{\alpha_{k}^{2}\hbar^{2}}{\left(2\pi\right)^{3}}\right)\left[\sum_{n}\frac{e_{n}^{2}}{R_{0n}^{10}}\left(2\int d\boldsymbol{x}\int d\boldsymbol{y}\frac{1}{|\boldsymbol{x}-\boldsymbol{y}|}\left(e^{-\frac{\boldsymbol{y}^{2}+\boldsymbol{x}^{2}}{2R_{0n}^{2}}}\left(\vec{\epsilon}_{\boldsymbol{k},\nu}\cdot\boldsymbol{y}\right)\left(\vec{\epsilon}_{\boldsymbol{k},\nu}\cdot\boldsymbol{x}\right)\right)\right)\right].\label{deriv A3_ad}
\end{equation}
The emission rate in Eq.~(\ref{ratedef_ad}) then becomes
\begin{equation}
\frac{d}{d\omega_k}\Gamma_{t}=\frac{k^{2}}{c}\sum_{\nu}\int d\Omega_{k}\frac{d}{dt}A_{3},\label{ratenewnew}
\end{equation}
and using 
\[
\int d\Omega_{k}\sum_{\nu}\epsilon_{\mathbf{k}\nu}^{i}\epsilon_{\mathbf{k}\nu}^{j}=\frac{8}{3}\pi\delta^{ij},
\]
we get
\[
\frac{d}{d\omega_k}\Gamma_{t}=\frac{k^{2}}{c}\frac{1}{\hbar^{2}\omega_{k}^{2}}\frac{4\pi G}{\hbar}\left(\frac{\alpha_{k}^{2}\hbar^{2}}{\left(2\pi\right)^{3}}\right)2\left(\frac{8}{3}\pi\right)\left[\sum_{n}\frac{e_{n}^{2}}{R_{0n}^{10}}\left(\int d\boldsymbol{x}\int d\boldsymbol{y}\frac{1}{|\boldsymbol{x}-\boldsymbol{y}|}e^{-\frac{\boldsymbol{y}^{2}+\boldsymbol{x}^{2}}{2R_{0n}^{2}}}\boldsymbol{y}\cdot\boldsymbol{x}\right)\right].
\]

We now focus on the integrals, which can be computed with the change of variable $\boldsymbol{s}=\boldsymbol{x}-\boldsymbol{y}$ and performing the integrations in polar coordinates. Then one obtains:
\begin{equation}
\int d\boldsymbol{x}\int d\boldsymbol{y}\frac{1}{|\boldsymbol{x}-\boldsymbol{y}|}e^{-\frac{\boldsymbol{y}^{2}+\boldsymbol{x}^{2}}{2R_{0n}^{2}}}\boldsymbol{y}\cdot\boldsymbol{x}=4\pi^{5/2}R_{0n}^{7}.
\end{equation}
Then, replacing $\alpha_{k}^{2}=\frac{\hbar}{2\varepsilon_{0}\omega_{k}(2\pi)^{3}}$,
we finally get the result: 
\begin{equation}
\frac{d}{d\omega_k}\Gamma_{t}=\frac{2G}{3\pi^{3/2}\varepsilon_{0}c^{3}\omega_{k}}\left[\sum_{n}\frac{e_{n}^{2}}{R_{0n}^{3}}\right].
\end{equation}
When we apply this result to an atom with atomic number $N$, there is a contribution due to the $N$ electrons with charge $e$ and one from the nucleus with charge $Ne$ leading to:
\begin{equation}\label{rateatom}
\frac{d}{d\omega_k}\Gamma_{t}=\frac{2Ge^2}{3\pi^{3/2}\varepsilon_{0}c^{3}\omega_{k}}\left[\frac{N}{R_{0e}^{3}}+\frac{{N}^{2}}{R_{0}^{3}}\right],
\end{equation}

In the case of electrons, $R_{0e}$ is at least of order of the Bohr radius while for the nuclei $R_{0}$ is of order of few femtometers. Therefore, the contribution from the nuclei is dominant and, considering a crystal with $N_{A}$ atoms, one gets:
\begin{equation}
\frac{d}{d\omega_k}\Gamma_{t}=\frac{2Ge^{2}N^{2}N_{A}}{3\pi^{3/2}\varepsilon_{0}c^{3}R_{0}^{3}\omega_{k}},
\end{equation}
which is precisely Eq.~(4) of the main text.

\subsection{Evaluation of $A_{1}$, $A_{2}$ and $A_{3}$.}\label{A1A2A3}

The interaction Hamiltonian, introduced in Eq. (15) of the Methods, is the sum of two terms:
\begin{equation}
H_{\text{\tiny INT}}(t)=H_{1}(t)+H_{2}(t),\label{splitHint_ad}
\end{equation}
both depending on $\boldsymbol{A}$ evolved in the (inverse) interaction
picture i.e.
\begin{equation}
\boldsymbol{A}(\boldsymbol{x},t)=\int d\boldsymbol{k}\sum_{\nu}\alpha_{k}\left[\vec{\epsilon}_{\boldsymbol{k},\nu}\,a_{\boldsymbol{k},\nu}\,e^{i\boldsymbol{k}\cdot\boldsymbol{x}+i\omega_{k}t}+\vec{\epsilon}_{\boldsymbol{k},\nu}\,a_{\boldsymbol{k},\nu}^{\dagger}\,e^{-i\boldsymbol{k}\cdot\boldsymbol{x}-i\omega_{k}t}\right].\label{Aint_ad}
\end{equation}
Explicitly they are: 
\begin{equation}
H_{1}(t):=\sum_{j}\left(-\frac{e_{j}}{m_{j}}\right)\boldsymbol{A}(\boldsymbol{x}_{j}(t),t)\cdot\boldsymbol{p}_{j}(t)=\int d\boldsymbol{k}\sum_{\nu}\left(R_{\boldsymbol{k},\nu}(t)\,a_{\boldsymbol{k},\nu}+R_{\boldsymbol{k},\nu}^{\dagger}(t)\,a_{\boldsymbol{k},\nu}^{\dagger}\right)\label{H1(t)_ad}
\end{equation}
with 
\begin{equation}
R_{\boldsymbol{k},\nu}(t):=-\alpha_{k}e^{+i\omega_{k}t}\sum_{j=1}^{N_{p}}\frac{e_{j}}{m_{j}}\vec{\epsilon}_{\boldsymbol{k},\nu}\cdot\boldsymbol{p}_{j}(t)e^{i\boldsymbol{k}\cdot\boldsymbol{x}_{j}(t)}\label{Rkmu_ad}
\end{equation}
and
\begin{equation}
H_{2}(t)=\sum_{j}\frac{e_{j}^{2}}{2m_{j}}\boldsymbol{A}^{2}(\boldsymbol{x}_{j}(t),t).\label{H2(t)_ad}
\end{equation}

\subsubsection*{Evaluation of $A_{1}$}

We start with evaluating $A_{1}$ of Eq.~(\ref{A1_ad}). As a first step
we need to compute
\begin{equation}
\mathcal{C}_{t}a_{\mathbf{k}\nu}^{\dagger}a_{\mathbf{k}\nu}=\frac{i}{\hbar}\left[H_{\text{\tiny INT}}(t),a_{\mathbf{k}\nu}^{\dagger}a_{\mathbf{k}\nu}\right]=a_{\mathbf{k}\nu}^{\dagger}\mathcal{C}_{t}a_{\mathbf{k}\nu}+\mathcal{C}_{t}[a_{\mathbf{k}\nu}^{\dagger}]a_{\mathbf{k}\nu}.\label{Cata_ad-1}
\end{equation}
The application of the superoperator $\mathcal{L}_{t_{1}}$ does not
affect the EM operators $a_{\mathbf{k}\nu}$ and $a_{\mathbf{k}\nu}^{\dagger}$,
therefore from Eqs.~(\ref{A1_ad}), (\ref{Cata_ad-1}) we see immediately
that $A_{1}$ is the sum of two terms, one having an $a_{\mathbf{k}\nu}^{\dagger}$
on the left and one having an $a_{\mathbf{k}\nu}$ on the right. When
the quantum average over the EM vacuum is taken, both contributions
are equal to zero, which implies $A_{1}=0$.

\subsubsection*{Evaluation of $A_{3}$}
We are left only with the two terms $A_{2}$ and $A_{3}$,
both containing twice the superoperator $\mathcal{C}_{t}$. Since we are
interested only in the contributions up to order $e^{2}$, we can neglect $H_{2}(t)$.
 
In the next subsection we show how the term $A_{2}$ gives no contribution. Therefore, we focus here on the relevant term $A_3$ defined in Eq.~(\ref{A3_ad}). Using the relations
\begin{equation}
\mathcal{C}_{t}a_{\mathbf{k}\nu}=\frac{i}{\hbar}\left[H_{1}(t),a_{\mathbf{k}\nu}\right]=-\frac{i}{\hbar}R_{\boldsymbol{k},\nu}^{\dagger}(t)\label{Ca_ad-2}
\end{equation}
and 
\begin{equation}
\mathcal{C}_{t}[a_{\mathbf{k}\nu}^{\dagger}]=\left(\mathcal{C}_{t}a_{\mathbf{k}\nu}\right)^{\dagger}=\frac{i}{\hbar}R_{\boldsymbol{k},\nu}(t),\label{Cat_ad-2}
\end{equation}
one obtains: 
\begin{equation}
\mathcal{C}_{t}a_{\mathbf{k}\nu}^{\dagger}a_{\mathbf{k}\nu}=-\frac{i}{\hbar}a_{\mathbf{k}\nu}^{\dagger}R_{\boldsymbol{k},\nu}^{\dagger}(t)+\frac{i}{\hbar}R_{\boldsymbol{k},\nu}(t)a_{\mathbf{k}\nu}.\label{Cata_ad-2}
\end{equation}
Then
\begin{equation}
\mathcal{C}_{t_{2}}\mathcal{C}_{t_{3}}a_{\mathbf{k}\nu}^{\dagger}a_{\mathbf{k}\nu}=\mathcal{C}_{t_{2}}\left(a_{\mathbf{k}\nu}^{\dagger}\mathcal{C}_{t_{3}}a_{\mathbf{k}\nu}+\mathcal{C}_{t_{3}}[a_{\mathbf{k}\nu}^{\dagger}]a_{\mathbf{k}\nu}\right)=\label{ccaa_ad}
\end{equation}
\[
=\mathcal{C}_{t_{2}}[a_{\mathbf{k}\nu}^{\dagger}]\mathcal{C}_{t_{3}}[a_{\mathbf{k}\nu}]+\mathcal{C}_{t_{3}}[a_{\mathbf{k}\nu}^{\dagger}]\mathcal{C}_{t_{2}}[a_{\mathbf{k}\nu}]=\frac{1}{\hbar^{2}}\left(R_{\boldsymbol{k},\nu}(t_{2})R_{\boldsymbol{k},\nu}^{\dagger}(t_{3})+R_{\boldsymbol{k},\nu}(t_{3})R_{\boldsymbol{k},\nu}^{\dagger}(t_{2})\right),
\]
where in the third step we neglected the terms that are null when
the average on the EM vacuum is taken. We make use of the fact
that this term enters in Eq.~(\ref{A3_ad}) inside a double integral
over $t_{2}$ and $t_{3}$, which implies:
\[
\int_{0}^{t_{1}}dt_{2}\int_{0}^{t_{2}}dt_{3}\mathcal{C}_{t_{2}}\mathcal{C}_{t_{3}}a_{\mathbf{k}\nu}^{\dagger}a_{\mathbf{k}\nu}=\frac{1}{\hbar^{2}}\int_{0}^{t_{1}}dt_{2}\int_{0}^{t_{2}}dt_{3}\left(R_{\boldsymbol{k},\nu}(t_{2})R_{\boldsymbol{k},\nu}^{\dagger}(t_{3})+R_{\boldsymbol{k},\nu}(t_{3})R_{\boldsymbol{k},\nu}^{\dagger}(t_{2})\right)=
\]
\[
\frac{1}{\hbar^{2}}\int_{0}^{t_{1}}dt_{2}\int_{0}^{t_{2}}dt_{3}R_{\boldsymbol{k},\nu}(t_{2})R_{\boldsymbol{k},\nu}^{\dagger}(t_{3})+\frac{1}{\hbar^{2}}\int_{0}^{t_{1}}dt_{3}\int_{t_{3}}^{t_{1}}dt_{2}R_{\boldsymbol{k},\nu}(t_{3})R_{\boldsymbol{k},\nu}^{\dagger}(t_{2})=
\]
\[
=\frac{1}{\hbar^{2}}\int_{0}^{t_{1}}dt_{2}\int_{0}^{t_{1}}dt_{3}R_{\boldsymbol{k},\nu}(t_{2})R_{\boldsymbol{k},\nu}^{\dagger}(t_{3}),
\]
where in the last step we just exchanged the variables of the second
integral $t_{2}\longleftrightarrow t_{3}$. 

Then, we can rewrite Eq.~(\ref{A3_ad}) as follows:
\begin{align}
A_{3}&=\int_{0}^{t}\!\!dt_{1}\int_{0}^{t_{1}}\!\!dt_{2}\int_{0}^{t_{1}}\!\!dt_{3}\textrm{Tr}\left[\rho_{g}\left(\frac{1}{\hbar^{2}}\int d\boldsymbol{Q}\sum_{n,n'}\tilde{\Gamma}_{n,n'}(\boldsymbol{Q})\left[e^{-\frac{i}{\hbar}\boldsymbol{Q}\cdot\boldsymbol{x}_{n'}(t_{1})}R_{\boldsymbol{k},\nu}(t_{2})R_{\boldsymbol{k},\nu}^{\dagger}(t_{3})e^{\frac{i}{\hbar}\boldsymbol{Q}\cdot\boldsymbol{x}_{n}(t_{1})}\right.\right.\right.\nonumber\\
&\left.\left.\left.-\frac{1}{2}\left\{ R_{\boldsymbol{k},\nu}(t_{2})R_{\boldsymbol{k},\nu}^{\dagger}(t_{3}),e^{-\frac{i}{\hbar}\boldsymbol{Q}\cdot(\boldsymbol{x}_{n'}(t_{1})-\boldsymbol{x}_{n}(t_{1}))}\right\} \right]\right)\right]\label{A3-2_ad}
\end{align}
where, using Eq.~(\ref{Rkmu_ad}),
\[
R_{\boldsymbol{k},\nu}(t_{2})R_{\boldsymbol{k},\nu}^{\dagger}(t_{3})=\alpha_{k}^{2}e^{+i\omega_{k}(t_{2}-t_{3})}\sum_{j,j'}\frac{e_{j}e_{j'}}{m_{j}m_{j'}}\left(\vec{\epsilon}_{\boldsymbol{k},\nu}\cdot\boldsymbol{p}_{j}(t_{2})e^{i\boldsymbol{k}\cdot\boldsymbol{x}_{j}(t_{2})}\right)\left(\vec{\epsilon}_{\boldsymbol{k},\nu}\cdot\boldsymbol{p}_{j'}(t_{3})e^{-i\boldsymbol{k}\cdot\boldsymbol{x}_{j'}(t_{3})}\right).
\]
We now focus on the first term of Eq.~(\ref{A3-2_ad}) and in particular
on
\begin{align}
T_{1}&:=\int_{0}^{t_{1}}dt_{2}\int_{0}^{t_{1}}dt_{3}\langle g|e^{-\frac{i}{\hbar}\boldsymbol{Q}\cdot\boldsymbol{x}_{n'}(t_{1})}R_{\boldsymbol{k},\nu}(t_{2})R_{\boldsymbol{k},\nu}^{\dagger}(t_{3})e^{\frac{i}{\hbar}\boldsymbol{Q}\cdot\boldsymbol{x}_{n}(t_{1})}|g\rangle=\nonumber\\
&=\sum_{f}\langle g|e^{-\frac{i}{\hbar}\boldsymbol{Q}\cdot\boldsymbol{x}_{n'}(t_{1})}\left(\int_{0}^{t_{1}}dt_{2}R_{\boldsymbol{k},\nu}(t_{2})\right)|f\rangle\langle f|\left(\int_{0}^{t_{1}}dt_{3}R_{\boldsymbol{k},\nu}^{\dagger}(t_{3})\right)e^{\frac{i}{\hbar}\boldsymbol{Q}\cdot\boldsymbol{x}_{n}(t_{1})}|g\rangle\label{T1_ad}
\end{align}
where the completeness relation $1=\sum_f|f\rangle\langle f|$ is written in terms of the free Hamiltonian eigenstates $|f\rangle$ with eigenvalues $E_f$. The first matrix element in Eq.~(\ref{T1_ad}) is    
\begin{align}
&\langle g|e^{-\frac{i}{\hbar}\boldsymbol{Q}\cdot\boldsymbol{x}_{n'}(t_{1})}\left(\int_{0}^{t_{1}}dt_{2}R_{\boldsymbol{k},\nu}(t_{2})\right)|f\rangle=\int_{0}^{t_{1}}dt_{2}\sum_{E}\langle g|e^{-\frac{i}{\hbar}\boldsymbol{Q}\cdot\boldsymbol{x}_{n'}(t_{1})}|E\rangle\langle E|R_{\boldsymbol{k},\nu}(t_{2})|f\rangle=\nonumber\\
&=\sum_{E}e^{-\frac{i}{\hbar}(E_{g}-E)t_{1}}\left(\int_{0}^{t_{1}}dt_{2}e^{-\frac{i}{\hbar}(E-E_{f}-\hbar\omega_{k})t_{2}}\right)\langle g|e^{-\frac{i}{\hbar}\boldsymbol{Q}\cdot\boldsymbol{x}_{n'}}|E\rangle\langle E|R_{\boldsymbol{k},\nu}(0)|f\rangle=\nonumber\\
&=\sum_{E}\left(\frac{e^{-\frac{i}{\hbar}(E_{g}-E_{f}-\hbar\omega_{k})t_{1}}-e^{-\frac{i}{\hbar}(E_{g}-E)t_{1}}}{-\frac{i}{\hbar}(E-E_{f}-\hbar\omega_{k})}\right)\langle g|e^{-\frac{i}{\hbar}\boldsymbol{Q}\cdot\boldsymbol{x}_{n'}}|E\rangle\langle E|R_{\boldsymbol{k},\nu}(0)|f\rangle.
\end{align}
The second term in the round bracket is a consequence of the implicit assumption that the interaction
is suddenly switch on at time $t=0$. This is
clearly not true for the problem we are studying, since both the
interaction with the EM vacuum as well as the gravitational related
collapse are always present. This term gives rise to unphysical contributions,
a problem which has been studied in detail in several articles~\cite{abd, bd, dirk, donadi2014emission}.
The simplest way to fix it is to assume an adiabatic switch
on of the potential, which is implemented by adding an exponential factor $e^{\epsilon t}$
with $\epsilon>0$ to the interaction, start the time integration
at minus infinity so that 
\begin{equation}
\int_{0}^{t_{1}}dt_{2}e^{-\frac{i}{\hbar}(E-E_{f}-\hbar\omega_{k})t_{2}}\;\;\longrightarrow\;\;\int_{-\infty}^{t_{1}}dt_{2}e^{-\frac{i}{\hbar}(E-E_{f}-\hbar\omega_{k})t_{2}+\epsilon t_{2}}=\frac{e^{\left[-\frac{i}{\hbar}(E-E_{f}-\hbar\omega_{k})+\epsilon\right]t_{1}}}{\epsilon-\frac{i}{\hbar}(E-E_{f}-\hbar\omega_{k})}\label{Inteps_ad}
\end{equation}
and take the limit $\epsilon\rightarrow0$ at the end of the calculation,
after all the time integrations or derivatives are performed.
Using Eq.~(\ref{Inteps_ad}), the matrix element we are computing
becomes: 
\begin{align}
&\langle g|e^{-\frac{i}{\hbar}\boldsymbol{Q}\cdot\boldsymbol{x}_{n'}(t_{1})}\!\int_{0}^{t_{1}}dt_{2}R_{\boldsymbol{k},\nu}(t_{2})|f\rangle\!=\!\sum_{E}\!\!\left(\frac{e^{\left[-\frac{i}{\hbar}(E_{g}-E_{f}-\hbar\omega_{k})+\epsilon\right]t_{1}}}{\epsilon-\frac{i}{\hbar}(E-E_{f}-\hbar\omega_{k})}\right)\!\!\langle g|e^{-\frac{i}{\hbar}\boldsymbol{Q}\cdot\boldsymbol{x}_{n'}}|E\rangle\langle E|R_{\boldsymbol{k},\nu}(0)|f\rangle\nonumber\\
&\simeq\frac{e^{\left[-\frac{i}{\hbar}(E_{g}-E_{f}-\hbar\omega_{k})+\epsilon\right]t_{1}}}{\epsilon+i\omega_{k}}\langle g|e^{-\frac{i}{\hbar}\boldsymbol{Q}\cdot\boldsymbol{x}_{n'}}R_{\boldsymbol{k},\nu}(0)|f\rangle,
\end{align}
where in the last step we used the fact that the typical binding energies
between electrons and nuclei are of the order 10 keV, those between
the nuclei in the crystal are even weaker and both are
much smaller than the energies of the emitted photons considered in the analysis (which are in the range $1000 \div 3800\; \mbox{keV}$). Going
back to Eq.~(\ref{T1_ad}) we find 
\begin{align}
T_{1}&=\sum_{f}\frac{e^{2\epsilon t_{1}}}{\epsilon^{2}+\omega_{k}^{2}}\langle g|e^{-\frac{i}{\hbar}\boldsymbol{Q}\cdot\boldsymbol{x}_{n'}}R_{\boldsymbol{k},\nu}(0)|f\rangle\langle f|R_{\boldsymbol{k},\nu}^{\dagger}(0)e^{\frac{i}{\hbar}\boldsymbol{Q}\cdot\boldsymbol{x}_{n}(t_{1})}|g\rangle\nonumber\\
&=\frac{e^{2\epsilon t_{1}}}{\epsilon^{2}+\omega_{k}^{2}}\langle g|e^{-\frac{i}{\hbar}\boldsymbol{Q}\cdot\boldsymbol{x}_{n'}}R_{\boldsymbol{k},\nu}(0)R_{\boldsymbol{k},\nu}^{\dagger}(0)e^{\frac{i}{\hbar}\boldsymbol{Q}\cdot\boldsymbol{x}_{n}(t_{1})}|g\rangle.
\end{align}
The same argument can be applied to the anticommutator in Eq.~(\ref{A3-2_ad})
and we will get the same factor. 

Since we are interested in computing the emission rate, we will need
\begin{equation}
\frac{d}{dt}A_{3}=\underset{\epsilon\rightarrow0}{\lim}\frac{d}{dt}\int_{-\infty}^{t}dt_{1}\frac{e^{3\epsilon t_{1}}}{\epsilon^{2}+\omega_{k}^{2}}\frac{1}{\hbar^{2}}\textrm{Tr}\left[\rho_{g}X\right]=\underset{\epsilon\rightarrow0}{\lim}\frac{e^{3\epsilon t}}{\epsilon^{2}+\omega_{k}^{2}}\frac{1}{\hbar^{2}}\textrm{Tr}\left[\rho_{g}X\right]=\frac{1}{\hbar^{2}\omega_{k}^{2}}\textrm{Tr}\left[\rho_{g}X\right],\label{derivative A3_ad}
\end{equation}
where we introduced 
\begin{equation}
X:=\!\!\int \!\!d\boldsymbol{Q}\sum_{n,n'}\tilde{\Gamma}_{n,n'}(\boldsymbol{Q})\!\left[e^{-\frac{i}{\hbar}\boldsymbol{Q}\cdot\boldsymbol{x}_{n'}}R_{\boldsymbol{k},\nu}(0)R_{\boldsymbol{k},\nu}^{\dagger}(0)e^{\frac{i}{\hbar}\boldsymbol{Q}\cdot\boldsymbol{x}_{n}}-\frac{1}{2}\left\{ R_{\boldsymbol{k},\nu}(0)R_{\boldsymbol{k},\nu}^{\dagger}(0),e^{-\frac{i}{\hbar}\boldsymbol{Q}\cdot(\boldsymbol{x}_{n'}-\boldsymbol{x}_{n})}\right\} \right]\!.\label{X_ad}
\end{equation}
In order to make explicit some symmetry properties of the calculation, it is convenient to rewrite $X$ by performing the
integration over $\boldsymbol{Q}$. After a straightforward
calculation (see also~\cite{bahrami2014role}), this gives:
\begin{align}
X & = \frac{8\pi G}{\hbar}\!\!\int\!\!\! d\boldsymbol{x}\int\!\!\! d\boldsymbol{y}\frac{1}{|\boldsymbol{x}-\boldsymbol{y}|}\!\left(\!M(\boldsymbol{x})R_{\boldsymbol{k},\nu}(0)R_{\boldsymbol{k},\nu}^{\dagger}(0)M(\boldsymbol{y})-\frac{1}{2}\left\{ M(\boldsymbol{x})M(\boldsymbol{y}),R_{\boldsymbol{k},\nu}(0)R_{\boldsymbol{k},\nu}^{\dagger}(0)\right\} \!\right)\nonumber\\
 & = -\frac{4\pi G}{\hbar}\int d\boldsymbol{x}\int d\boldsymbol{y}\frac{1}{|\boldsymbol{x}-\boldsymbol{y}|}\left[M(\boldsymbol{y}),\left[M(\boldsymbol{x}),R_{\boldsymbol{k},\nu}(0)R_{\boldsymbol{k},\nu}^{\dagger}(0)\right]\right]\label{Xandrea_ad}
\end{align}
where
\begin{equation}
M(\boldsymbol{x})=\sum_{n}\mu_{n}(\boldsymbol{x},\boldsymbol{x}_n)=\sum_{n}m_{n}\frac{1}{\left(2\pi R_{0n}^{2}\right)^{\frac{3}{2}}}e^{-\frac{(\boldsymbol{x}-\boldsymbol{x}_{n})^{2}}{2R_{0n}^{2}}}:=\sum_{n}m_{n}g_n(\boldsymbol{x}).\label{M(x)_ad}
\end{equation}
$R_{0n}$ represents the spatial extension of the mass density of
the $n$-th particle (since we are considering both
electrons and nuclei, it is important to consider distinct values for 
 $R_{0}$). Writing explicitly the
``$R$'' terms we get 
\begin{align}
R_{\boldsymbol{k},\nu}(0)R_{\boldsymbol{k},\nu}^{\dagger}(0)&=\alpha_{k}^{2}\sum_{j,j'=1}^{N_{p}}\frac{e_{j}e_{j'}}{m_{j}m_{j'}}\left(\vec{\epsilon}_{\boldsymbol{k},\nu}\cdot\boldsymbol{p}_{j}e^{i\boldsymbol{k}\cdot\boldsymbol{x}_{j}}\right)\left(\vec{\epsilon}_{\boldsymbol{k},\nu}\cdot\boldsymbol{p}_{j'}e^{-i\boldsymbol{k}\cdot\boldsymbol{x}_{j'}}\right)\nonumber\\
&=\alpha_{k}^{2}\sum_{j,j'=1}^{N_{p}}\frac{e_{j}e_{j'}}{m_{j}m_{j'}}e^{i\boldsymbol{k}\cdot(\boldsymbol{x}_{j}-\boldsymbol{x}_{j'})}\left(\vec{\epsilon}_{\boldsymbol{k},\nu}\cdot\boldsymbol{p}_{j}\right)\left(\vec{\epsilon}_{\boldsymbol{k},\nu}\cdot\boldsymbol{p}_{j'}\right),\label{RRt(0)_ad}
\end{align}
where in the second step we used the fact that we are working in Coulomb
gauge, so $\vec{\epsilon}_{\boldsymbol{k},\nu}\cdot\boldsymbol{k}=0$.
Then
\[
X=-\frac{4\pi G \alpha_{k}^{2}}{\hbar}\sum_{n,n',j,j'}\!\!\frac{m_{n'}m_{n}e_{j}e_{j'}}{m_{j}m_{j'}}\!\!\!\int \!\!d\boldsymbol{x}\!\!\int \!\!d\boldsymbol{y}\frac{\left[g_{n'}(\boldsymbol{y}),\left[g_{n}(\boldsymbol{x}),e^{i\boldsymbol{k}\cdot(\boldsymbol{x}_{j}-\boldsymbol{x}_{j'})}\left(\vec{\epsilon}_{\boldsymbol{k},\nu}\cdot\boldsymbol{p}_{j}\right)\left(\vec{\epsilon}_{\boldsymbol{k},\nu}\cdot\boldsymbol{p}_{j'}\right)\right]\right]}{|\boldsymbol{x}-\boldsymbol{y}|}
\]
\begin{equation}
=-\frac{4\pi G \alpha_{k}^{2}}{\hbar}\sum_{n,n',j,j'}\!\!\frac{m_{n'}m_{n}e_{j}e_{j'}}{m_{j}m_{j'}}e^{i\boldsymbol{k}\cdot(\boldsymbol{x}_{j}-\boldsymbol{x}_{j'})}\!\!\!\int \!\!d\boldsymbol{x}\!\!\int \!\!d\boldsymbol{y}\frac{\left[g_{n'}(\boldsymbol{y}),\left[g_{n}(\boldsymbol{x}),\left(\vec{\epsilon}_{\boldsymbol{k},\nu}\cdot\boldsymbol{p}_{j}\right)\left(\vec{\epsilon}_{\boldsymbol{k},\nu}\cdot\boldsymbol{p}_{j'}\right)\right]\right]}{|\boldsymbol{x}-\boldsymbol{y}|}.\label{X2_ad}
\end{equation}
We now focus on the double commutator:
\begin{align}
C&:=\left[g_{n'}(\boldsymbol{y}),\left[g_{n}(\boldsymbol{x}),\left(\vec{\epsilon}_{\boldsymbol{k},\nu}\cdot\boldsymbol{p}_{j}\right)\left(\vec{\epsilon}_{\boldsymbol{k},\nu}\cdot\boldsymbol{p}_{j'}\right)\right]\right]\\
&=\left[g_{n'}(\boldsymbol{y}),\left(\vec{\epsilon}_{\boldsymbol{k},\nu}\cdot\boldsymbol{p}_{j}\right)\right]\left[g_{n}(\boldsymbol{x}),\left(\vec{\epsilon}_{\boldsymbol{k},\nu}\cdot\boldsymbol{p}_{j'}\right)\right]+\left[g_{n}(\boldsymbol{x}),\left(\vec{\epsilon}_{\boldsymbol{k},\nu}\cdot\boldsymbol{p}_{j}\right)\right]\left[g_{n'}(\boldsymbol{y}),\left(\vec{\epsilon}_{\boldsymbol{k},\nu}\cdot\boldsymbol{p}_{j'}\right)\right]\nonumber\label{C_ad-1}
\end{align}
where we used the fact that each commutator is just a function of position operators, indeed:
\begin{eqnarray}
\left[g_{n}(\boldsymbol{x}),\left(\vec{\epsilon}_{\boldsymbol{k},\nu}\cdot\boldsymbol{p}_{j}\right)\right]&=&\frac{\vec{\epsilon}_{\boldsymbol{k},\nu}}{\left(2\pi R_{0n}^{2}\right)^{\frac{3}{2}}}\cdot\left[e^{-\frac{(\boldsymbol{x}-\boldsymbol{x}_{n})^{2}}{2R_{0n}^{2}}},\boldsymbol{p}_{j}\right]=\frac{\vec{\epsilon}_{\boldsymbol{k},\nu}}{\left(2\pi R_{0n}^{2}\right)^{\frac{3}{2}}}\cdot\left(i\hbar\nabla_{j}e^{-\frac{(\boldsymbol{x}-\boldsymbol{x}_{n})^{2}}{2R_{0n}^{2}}}\right)\nonumber\\
&=&\delta_{jn}i\hbar\left(\vec{\epsilon}_{\boldsymbol{k},\nu}\cdot(\boldsymbol{x}-\boldsymbol{x}_{n})\right)\frac{e^{-\frac{(\boldsymbol{x}-\boldsymbol{x}_{n})^{2}}{2R_{0n}^{2}}}}{\left(2\pi\right)^{3/2}R_{0n}^{5}}.\label{mup_ad}
\end{eqnarray}
Then $C$ becomes
\begin{equation}
C=-\frac{\hbar^{2}\left(\delta_{jn'}\delta_{j'n}+\delta_{jn}\delta_{j'n'}\right)}{\left(2\pi\right)^{3}R_{0n'}^{5}R_{0n}^{5}}\left(e^{-\frac{(\boldsymbol{y}-\boldsymbol{x}_{n'})^{2}}{2R_{0n'}^{2}}-\frac{(\boldsymbol{x}-\boldsymbol{x}_{n})^{2}}{2R_{0n}^{2}}}\left(\vec{\epsilon}_{\boldsymbol{k},\nu}\cdot(\boldsymbol{x}-\boldsymbol{x}_{n})\right)\left(\vec{\epsilon}_{\boldsymbol{k},\nu}\cdot(\boldsymbol{y}-\boldsymbol{x}_{n'})\right)\right)\label{Cfinal_ad}
\end{equation}
and inserting it in Eq.~(\ref{X2_ad}) we get 
\begin{align}\label{tititi}
X&=\frac{4\pi G \alpha_{k}^{2}}{\hbar}\left(\frac{\hbar^{2}}{\left(2\pi\right)^{3}}\right)\sum_{n,n'}\frac{e_{n'}e_{n}}{R_{0n'}^{5}R_{0n}^{5}}\left\{ 2\cos\left[\boldsymbol{k}\cdot(\boldsymbol{x}_{n}-\boldsymbol{x}_{n'})\right]\times\right. \nonumber\\
&\left. \times\int d\boldsymbol{x}\int d\boldsymbol{y}\frac{1}{|\boldsymbol{x}-\boldsymbol{y}|}\left[e^{-\frac{(\boldsymbol{y}-\boldsymbol{x}_{n'})^{2}}{2R_{0n'}^{2}}-\frac{(\boldsymbol{x}-\boldsymbol{x}_{n})^{2}}{2R_{0n}^{2}}}\left(\vec{\epsilon}_{\boldsymbol{k},\nu}\cdot(\boldsymbol{x}-\boldsymbol{x}_{n})\right)\left(\vec{\epsilon}_{\boldsymbol{k},\nu}\cdot(\boldsymbol{y}-\boldsymbol{x}_{n'})\right)\right]\right\} .
\end{align}
Our goal is to compute $\textrm{Tr}\left[\rho_{g}X\right]$ and substitute it in Eq.~(\ref{derivative A3_ad}). Since $\rho_g$ describes the state of the crystal, where the distances between the nuclei and the electrons are much larger than the photons wavelengths we
are measuring (which are in the range $[10^{-4},\,10^{-3}]$ nm),
this implies $\boldsymbol{k}\cdot(\boldsymbol{x}_{n}-\boldsymbol{x}_{n'})\gg1$
for $n\neq n'$. Such contributions, when the sum over the different
directions of $\boldsymbol{k}$ is taken, averages to zero. Then, the only relevant terms are those with
$n=n'$ and, by taking the trace in the position basis (so that the position operators $\boldsymbol{x}_{n}$ in Eq.~(\ref{tititi}) become simple integration variables), performing the change of variables $\boldsymbol{x}-\boldsymbol{x}_{n}\rightarrow\boldsymbol{x}$ and $\boldsymbol{y}-\boldsymbol{x}_{n}\rightarrow\boldsymbol{y}$ one gets:
\[
\textrm{Tr}\left[\rho_{g}X\right]\simeq\frac{4\pi G}{\hbar}\left(\frac{\alpha_{k}^{2}\hbar^{2}}{\left(2\pi\right)^{3}}\right)\left[\sum_{n}\frac{e_{n}^{2}}{R_{0n}^{10}}\left(2\int d\boldsymbol{x}\int d\boldsymbol{y}\frac{1}{|\boldsymbol{x}-\boldsymbol{y}|}\left(e^{-\frac{\boldsymbol{y}^{2}+\boldsymbol{x}^{2}}{2R_{0n}^{2}}}\left(\vec{\epsilon}_{\boldsymbol{k},\nu}\cdot\boldsymbol{y}\right)\left(\vec{\epsilon}_{\boldsymbol{k},\nu}\cdot\boldsymbol{x}\right)\right)\right)\right].
\]
where we used the fact that, after the change of variables, the trace gives  $\textrm{Tr}\left[\rho_{g}\right]=1$. We finally get:
\[
\frac{d}{dt}A_{3}=\frac{1}{\hbar^{2}\omega_{k}^{2}}\frac{4\pi G}{\hbar}\left(\frac{\alpha_{k}^{2}\hbar^{2}}{\left(2\pi\right)^{3}}\right)\left[\sum_{n}\frac{e_{n}^{2}}{R_{0n}^{10}}\left(2\int d\boldsymbol{x}\int d\boldsymbol{y}\frac{1}{|\boldsymbol{x}-\boldsymbol{y}|}\left(e^{-\frac{\boldsymbol{y}^{2}+\boldsymbol{x}^{2}}{2R_{0n}^{2}}}\left(\vec{\epsilon}_{\boldsymbol{k},\nu}\cdot\boldsymbol{y}\right)\left(\vec{\epsilon}_{\boldsymbol{k},\nu}\cdot\boldsymbol{x}\right)\right)\right)\right]
\]
which is precisely Eq.~(\ref{deriv A3_ad}). 

\subsubsection*{Evaluation of $A_{2}$}\label{secA2}
In this section we compute the $A_{2}$ term in Eq.~(\ref{A2_ad}):
\begin{equation}
A_{2}:=\int_{0}^{t}dt_{1}\int_{0}^{t_{1}}dt_{2}\int_{0}^{t_{2}}dt_{3}\textrm{Tr}\left[\rho_{i}\mathcal{C}_{t_{1}}\mathcal{L}_{t_{2}}\mathcal{C}_{t_{3}}a_{\mathbf{k}\nu}^{\dagger}a_{\mathbf{k}\nu}\right]\!.\label{A2_ad-1}
\end{equation}
It will be shown that $A_{2}$ gives no contribution to the emission rate.
We focus on the action of the superoperators $\mathcal{C}$
and $\mathcal{L}$ on $a_{\mathbf{k}\nu}^{\dagger}a_{\mathbf{k}\nu}$.
Using Eqs.~(\ref{Ca_ad-2}), (\ref{Cat_ad-2}) and (\ref{Cata_ad-2}) one gets:
\begin{equation}
\mathcal{C}_{t_{1}}\mathcal{L}_{t_{2}}\mathcal{C}_{t_{3}}a_{\mathbf{k}\nu}^{\dagger}a_{\mathbf{k}\nu}=-\frac{i}{\hbar}\mathcal{C}_{t_{1}}\left(a_{\mathbf{k}\nu}^{\dagger}\mathcal{L}_{t_{2}}[R_{\boldsymbol{k},\nu}^{\dagger}(t_{3})]\right)+\frac{i}{\hbar}\mathcal{C}_{t_{1}}\left(\mathcal{L}_{t_{2}}[R_{\boldsymbol{k},\nu}(t_{3})]a_{\mathbf{k}\nu}\right)\label{CLC_ad}
\end{equation}
where we used the fact that $\mathcal{L}_{t_{2}}$ does not affect
$a_{\mathbf{k}\nu}$ and $a_{\mathbf{k}\nu}^{\dagger}$. Using the
relations
\begin{equation}
\mathcal{L}_{t}O^{\dagger}=[\mathcal{L}_{t}O]^{\dagger},
\end{equation}
\[
\mathcal{C}_{t}O^{\dagger}=\frac{i}{\hbar}\left[H_{1}(t),O^{\dagger}\right]=\left(-\frac{i}{\hbar}\left[O,H_{1}(t)\right]\right)^{\dagger}=\left(\mathcal{C}_{t}O\right)^{\dagger},
\]
Eq.~(\ref{CLC_ad}) can be written as follows
\begin{equation}\label{tutu}
\mathcal{C}_{t_{1}}\mathcal{L}_{t_{2}}\mathcal{C}_{t_{3}}a_{\mathbf{k}\nu}^{\dagger}a_{\mathbf{k}\nu}=\frac{i}{\hbar}\mathcal{C}_{t_{1}}\left(\mathcal{L}_{t_{2}}[R_{\boldsymbol{k},\nu}(t_{3})]a_{\mathbf{k}\nu}\right)+H.c.\;,
\end{equation}
where ``$H.c.$" stands for ``Hermitian conjugate". The first term of Eq.~(\ref{tutu}) is
\begin{equation}
\mathcal{C}_{t_{1}}\left(\mathcal{L}_{t_{2}}[R_{\boldsymbol{k},\nu}(t_{3})]a_{\mathbf{k}\nu}\right)=\mathcal{C}_{t_{1}}\left(\mathcal{L}_{t_{2}}[R_{\boldsymbol{k},\nu}(t_{3})]\right)a_{\mathbf{k}\nu}+\mathcal{L}_{t_{2}}[R_{\boldsymbol{k},\nu}(t_{3})]\left(-\frac{i}{\hbar}R_{k,\nu}^{\dagger}(t_{1})\right).\label{c1l2_ad}
\end{equation}
The first term of Eq.~(\ref{c1l2_ad}) gives a zero contribution when
the average on the EM vacuum is taken. Then, writing explicitly the
last term, we get
\begin{align}
\mathcal{C}_{t_{1}}\mathcal{L}_{t_{2}}\mathcal{C}_{t_{3}}a_{\mathbf{k}\nu}^{\dagger}a_{\mathbf{k}\nu}&=\frac{1}{\hbar^{2}}\int d\boldsymbol{Q}\sum_{n,n'}\tilde{\Gamma}_{n,n'}(\boldsymbol{Q})\left(e^{-\frac{i}{\hbar}\boldsymbol{Q}\cdot\boldsymbol{x}_{n'}(t_{2})}R_{\boldsymbol{k},\nu}(t_{3})e^{\frac{i}{\hbar}\boldsymbol{Q}\cdot\boldsymbol{x}_{n}(t_{2})}\right.\nonumber\\
&\left.-\frac{1}{2}\left\{ R_{\boldsymbol{k},\nu}(t_{3}),e^{-\frac{i}{\hbar}\boldsymbol{Q}\cdot(\boldsymbol{x}_{n'}(t_{2})-\boldsymbol{x}_{n}(t_{2}))}\right\} \right)\left(R_{\boldsymbol{k},\nu}^{\dagger}(t_{1})\right)+H.c.\;.
\end{align}
Going back to $A_{2}$ and taking on its time derivative, we have
\begin{align}
\frac{d}{dt}A_{2}&:=\frac{1}{\hbar^{2}}\int d\boldsymbol{Q}\sum_{n,n'}\tilde{\Gamma}_{n,n'}(\boldsymbol{Q})\int_{0}^{t}dt_{2}\int_{0}^{t_{2}}dt_{3}\langle g|\left(e^{-\frac{i}{\hbar}\boldsymbol{Q}\cdot\boldsymbol{x}_{n'}(t_{2})}R_{\boldsymbol{k},\nu}(t_{3})e^{\frac{i}{\hbar}\boldsymbol{Q}\cdot\boldsymbol{x}_{n}(t_{2})}\right.\nonumber\\
&\left.-\frac{1}{2}\left\{ R_{\boldsymbol{k},\nu}(t_{3}),e^{-\frac{i}{\hbar}\boldsymbol{Q}\cdot(\boldsymbol{x}_{n'}(t_{2})-\boldsymbol{x}_{n}(t_{2}))}\right\} \right)\left(R_{\boldsymbol{k},\nu}^{\dagger}(t)\right)|g\rangle+H.c.\;.\label{derivative a2_ad}
\end{align}
We consider the first term in the round bracket and we follow the
same steps used for computing $\frac{d}{dt}A_{3}$: we insert a completeness
in the energy eigenstates of the system and also the adiabatic switch on
of the interactions:
\begin{align}
&\int_{-\infty}^{t}\!\!\!dt_{2}e^{\epsilon t_{2}}\int_{-\infty}^{t_{2}}\!\!\!dt_{3}e^{\epsilon t_{3}}\!\!\!\sum_{{f},E,E'}\langle g|e^{-\frac{i}{\hbar}\boldsymbol{Q}\cdot\boldsymbol{x}_{n'}(t_{2})}|E'\rangle\langle E'|R_{\boldsymbol{k},\nu}(t_{3})|E\rangle\langle E|e^{\frac{i}{\hbar}\boldsymbol{Q}\cdot\boldsymbol{x}_{n}(t_{2})}|{f}\rangle\langle {f}|R_{\boldsymbol{k},\nu}^{\dagger}(t)|g\rangle=\nonumber\\
&=\sum_{f,E,E'}\int_{-\infty}^{t}dt_{2}e^{\left[-\frac{i}{\hbar}(E_{g}-E'+E-E_{f})t_{2}+\epsilon t_{2}\right]}\int_{-\infty}^{t_{2}}dt_{3}e^{\left[-\frac{i}{\hbar}(E'-E-\hbar\omega_{k})t_{3}+\epsilon t_{3}\right]}e^{-\frac{i}{\hbar}(E_{f}+\hbar\omega_{k}-E_{g})t}\times\nonumber\\
&\times\langle g|e^{-\frac{i}{\hbar}\boldsymbol{Q}\cdot\boldsymbol{x}_{n'}}|E'\rangle\langle E'|R_{\boldsymbol{k},\nu}(0)|E\rangle\langle E|e^{\frac{i}{\hbar}\boldsymbol{Q}\cdot\boldsymbol{x}_{n}}|{f}\rangle\langle {f}|R_{\boldsymbol{k},\nu}^{\dagger}(0)|g\rangle=\nonumber\\
&=-\sum_{{f},E,E'}\frac{\hbar^{2}e^{2t\epsilon}}{\left[E-E'+\hbar(\omega_{k}-i\epsilon)\right]\left[E_{f}-E_{g}+\hbar(\omega_{k}-2i\epsilon)\right]}\times \nonumber\\
&\times\langle g|e^{-\frac{i}{\hbar}\boldsymbol{Q}\cdot\boldsymbol{x}_{n'}}|E'\rangle\langle E'|R_{\boldsymbol{k},\nu}(0)|E\rangle\langle E|e^{\frac{i}{\hbar}\boldsymbol{Q}\cdot\boldsymbol{x}_{n}}|{f}\rangle\langle {f}|R_{\boldsymbol{k},\nu}^{\dagger}(0)|g\rangle\simeq\nonumber\\
&\simeq-\frac{e^{2t\epsilon}}{\left[(\omega_{k}-i\epsilon)\right]\left[(\omega_{k}-2i\epsilon)\right]}\langle g|e^{-\frac{i}{\hbar}\boldsymbol{Q}\cdot\boldsymbol{x}_{n'}}R_{\boldsymbol{k},\nu}(0)e^{\frac{i}{\hbar}\boldsymbol{Q}\cdot\boldsymbol{x}_{n}}R_{\boldsymbol{k},\nu}^{\dagger}(0)|g\rangle \rightarrow \nonumber\\
&\underset{\epsilon\rightarrow0}{\longrightarrow}\;\;-\frac{1}{\omega_{k}^{2}}\langle g|e^{-\frac{i}{\hbar}\boldsymbol{Q}\cdot\boldsymbol{x}_{n'}}R_{\boldsymbol{k},\nu}(0)e^{\frac{i}{\hbar}\boldsymbol{Q}\cdot\boldsymbol{x}_{n}}R_{\boldsymbol{k},\nu}^{\dagger}(0)|g\rangle.
\end{align}
Proceeding in the same way for the terms in the anticommutator of
Eq.~(\ref{derivative a2_ad}), we get
\begin{align}
\frac{d}{dt}A_{2}&:=-\frac{1}{\hbar^{2}\omega_{k}^{2}}\int d\boldsymbol{Q}\sum_{n,n'}\tilde{\Gamma}_{n,n'}(\boldsymbol{Q})\times \\
&\times \langle g|\left(e^{-\frac{i}{\hbar}\boldsymbol{Q}\cdot\boldsymbol{x}_{n'}}R_{\boldsymbol{k},\nu}(0)e^{\frac{i}{\hbar}\boldsymbol{Q}\cdot\boldsymbol{x}_{n}}-\frac{1}{2}\left\{ R_{\boldsymbol{k},\nu}(0),e^{-\frac{i}{\hbar}\boldsymbol{Q}\cdot(\boldsymbol{x}_{n'}-\boldsymbol{x}_{n})}\right\} \right)R_{\boldsymbol{k},\nu}^{\dagger}(0)|g\rangle+H.c.\;\nonumber.
\end{align}
Performing the integration over $\boldsymbol{Q}$, we get
\begin{align}
&\int d\boldsymbol{Q}\sum_{n,n'}\tilde{\Gamma}_{n,n'}(\boldsymbol{Q})\left(e^{-\frac{i}{\hbar}\boldsymbol{Q}\cdot\boldsymbol{x}_{n'}}R_{\boldsymbol{k},\nu}(0)e^{\frac{i}{\hbar}\boldsymbol{Q}\cdot\boldsymbol{x}_{n}}-\frac{1}{2}\left\{ R_{\boldsymbol{k},\nu}(0),e^{-\frac{i}{\hbar}\boldsymbol{Q}\cdot(\boldsymbol{x}_{n'}-\boldsymbol{x}_{n})}\right\} \right)\nonumber\\
&=-\frac{4\pi G}{\hbar}\int d\boldsymbol{x}\int d\boldsymbol{y}\frac{1}{|\boldsymbol{x}-\boldsymbol{y}|}\left[M(\boldsymbol{y}),\left[M(\boldsymbol{x}),R_{\boldsymbol{k},\nu}(0)\right]\right]
\end{align}
with $M(\boldsymbol{x})$ defined in Eq.~(\ref{M(x)_ad}) and $R_{\boldsymbol{k},\nu}(0)$
in Eq.~(\ref{Rkmu_ad}) setting $t=0$. Since $R_{\boldsymbol{k},\nu}(0)$
is linear in the momentum operator, the first commutator will result in a function of position
operators only (see Eq.~(\ref{mup_ad}))
which implies that the second commutation gives zero. Therefore
\begin{equation}
\frac{d}{dt}A_{2}=0.
\end{equation}
This concludes the calculations.

\end{document}